\documentclass[journal]{IEEEtran}
\IEEEoverridecommandlockouts
\pdfoutput=1
\usepackage{cite}
\usepackage{soul}
\usepackage{amsmath,amssymb,amsfonts}
\usepackage{algorithmic}
\usepackage{graphicx}
\usepackage{textcomp}
\usepackage{xcolor}
\usepackage{float}
\usepackage{array}
\usepackage{stfloats}
\usepackage{algorithmic}
\usepackage{gensymb}
\usepackage{caption}
\usepackage[ruled,vlined]{algorithm2e}
\usepackage{subcaption}
\usepackage{multirow}
\usepackage{hyperref}
\usepackage{makecell}
\hypersetup{pdfcreator={Me}}
\hypersetup{
    colorlinks=true,
    linkcolor=blue,     
    urlcolor=cyan,
    }

\usepackage{pgfplots}
\pgfplotsset{width=5.5cm,compat=1.9}
\usepgfplotslibrary{external}
\begin{document}
\title{ASL Trigger Recognition in Mixed Activity/Signing Sequences for RF Sensor-Based User Interfaces}

\author{Emre~Kurto\u{g}lu,~\IEEEmembership{Student Member,~IEEE,}
        Ali C.~Gurbuz,~\IEEEmembership{Senior Member, IEEE,}
        Evie A. Malaia, \\ Darrin Griffin, Chris Crawford,
        and~Sevgi Z. ~Gurbuz,~\IEEEmembership{Senior Member,~IEEE}
        }

\maketitle
\begin{abstract}
The past decade has seen great advancements in speech recognition for control of interactive devices, personal assistants, and computer interfaces.  However, Deaf and hard-of-hearing (HoH) individuals, whose primary mode of communication is sign language, cannot use voice-controlled interfaces.  Although there has been significant work in video-based sign language recognition, video is not effective in the dark and has raised privacy concerns in the Deaf community when used in the context of human ambient intelligence.  RF sensors have been recently proposed as a new modality that can be effective under the circumstances where video is not.  This paper considers the problem of recognizing a trigger sign (wake word) in the context of daily living, where gross motor activities are interwoven with signing sequences.  The proposed approach exploits multiple RF data domain representations (time-frequency, range-Doppler, and range-angle) for sequential classification of mixed motion data streams. The recognition accuracy of signs with varying kinematic properties is compared and used to make recommendations on appropriate trigger sign selection for RF-sensor based user interfaces.  The proposed approach achieves a trigger sign detection rate of 98.9\% and a classification accuracy of 92\%  for 15 ASL words and 3 gross motor activities. 
\end{abstract}


\IEEEpeerreviewmaketitle

\section{Introduction}
\label{sec:Intro}
\IEEEPARstart{T}{he} past decade has seen great advancements in sensing for ambient intelligence, including speech recognition for control of interactive devices, personal assistants, and human-computer interfaces.  However, deaf and hard-of-hearing (HoH) individuals, whose primary mode of communication is sign language, cannot benefit from voice-controlled interfaces.  Research in recognition of American Sign Language (ASL) has focused primarily on wearable sensors \cite{Sun2016_IMU_sEMG,Siddiqui2021,6814287}, optical cameras \cite{6814287,Koller2020_ASLvideo,Cui2019_ASLvideo}, and infrared depth sensors \cite{Sun2013_ASLkinect,Mittal2019_ASLleap}.  Wearables, such as ``signing'' gloves embedded with inertial measurement units (IMUs), or surface elecromyography (sEMG) sensors have yielded relatively higher recognition accuracy, but inhibit natural motion, and are thus not highly preferred by members of the Deaf community \cite{bragg2019sign}.

Video cameras are perhaps the most often used device by deaf/HoH individuals for interpersonal communications.  RGB-D cameras, which add depth measurements to video recordings for the purposes of skeleton tracking, such as Kinect or Leap Motion sensors, have improved recognition accuracy relative to video-only approaches.  However, RGB-D cameras are ineffective under low illumination and may invade privacy through the acquisition of personal imagery of the face and environment.

RF sensors have been recently proposed \cite{ASL_Patent2018,Gurbuz_ASLR_RadarCon,Gurbuz2020_ASL_SenConf,Gurbuz2021_ASLR} as a new modality for ASL recognition that has the capability of measurning human kinematics through fine range, angle and velocity measurements.  RF sensors are also effective in the dark and do not make any visual recordings of the people or environment.
RF sensors, also known as radar, which is short for \textit{radio detection and ranging}, acquire independent measurements of distance, velocity, and angle.  Using a technique known as \textit{stretch processing} \cite{Gurbuz_DNN4Radar2020}, the frequency difference between the transmitted and received frequency modulated continuous wave (FMCW) signals can be used to compute the round-trip travel time of the signal, and, hence, distance to an object. The Doppler shift, on the other hand, relates radial velocity to the frequency shift in the received signal.  Rotations or vibrations result in additional Doppler frequency modulations, known as micro-Doppler ($\mu D$) frequencies, centered around the main Doppler shift due to translational motion \cite{9106153}.  The $\mu D$ signature is a 2D time-frequency representation of the RF data, which reveals the unique kinematics of the observed motion.  Thus, $\mu D$ has been exploited as a biometric \cite{1550191} for recognizing individuals \cite{8333730}, activities \cite{4801689}, aided/unaided walking \cite{szgMicroDoppler}, falls \cite{7426551,6945894} and even different gaits \cite{8613848}.

Although RF sensors cannot effectively perceive facial expressions or hand shape, radar does provide data that is complementary to that of video:  while video is effective in capturing spatial parameters, radar is more adept at capturing temporal or dynamic parameters.  This is because radar measurements of distance and velocity are based on independent physics-based measurements:  distance is computed from round-trip travel time, while velocity is computed from the Doppler shift, which has greater accuracy than the computation of displacement over time.  Thus, although RF sensors cannot be used to reconstruct facial expressions or hand shape, radar can provide a new way of recognizing signs in a non-contact, ambient fashion based primarily on signing kinematics and range profiles.   

As a result, there has been much research on the use of RF sensing for hand gesture recognition \cite{Gesture_Gu2019,GestureRadar2021}, especially since the development of low-cost, low-power, high resolution, integrated millimeter wave RF transceivers \cite{Arbabian2012}. However, most current research involves controlled data acquisition with the participant located in a fixed position relative to the radar, articulating only a single gesture or sign. A critical challenge that has not been adequately addressed in the literature, however, is the challenge of ASL recognition in the context of daily living. To the best of our knowledge, this work represents the first to consider triggering and command recognition of RF-sensor enabled devices under more realistic conditions, where the RF data is acquired in a continuous fashion to capture mixed sequences of gross body motion/activity intertwined with ASL signing.

\begin{figure*}[t!]
\centering
\includegraphics[width=17 cm]{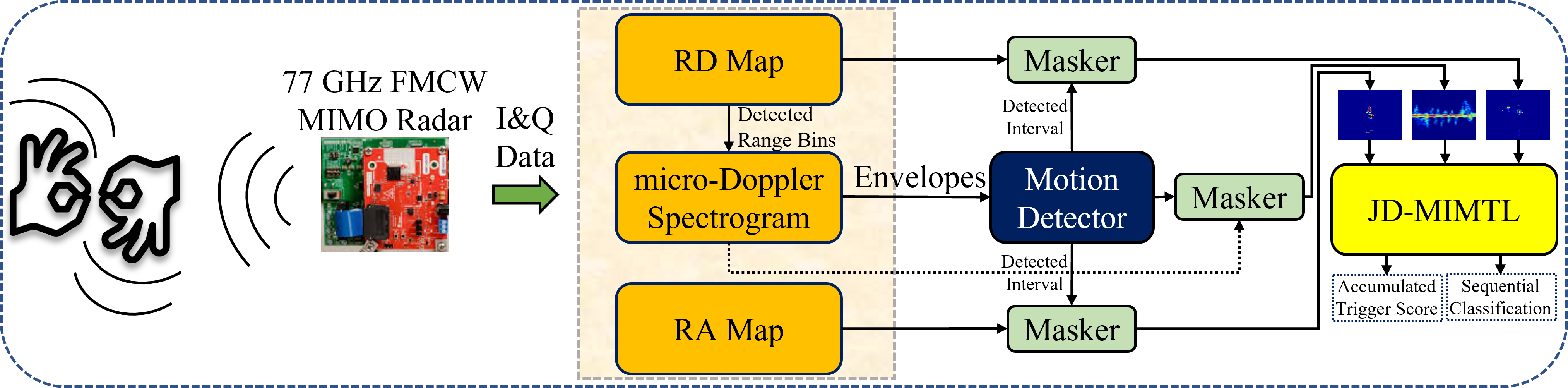}
  \centering \caption{Flowchart for the proposed approach.}
  \label{fig:flow_diagram}
\end{figure*}

 In particular, we analyze the design considerations for selection of a trigger sign based on kinematics, replicability, and recognition accuracy.  Whereas current approaches rely on just one RF data representation, we propose a joint-domain, multi-input, multi-task learning (JD-MIMTL) framework coupled with a motion detector to isolate the intervals over which the user is engaged in meaningful movement, and thus prevent unnecessary expenditure of computation resources when the RF system is not being used. Figure \ref{fig:flow_diagram} shows a flowchart providing an overview of the proposed approach. Our results show that the proposed approach exceeds that offered by approaches common in the literature and can recognize a sequence of 3 activities and 15 ASL signs with 92\% accuracy, while detecting trigger signs with rates as high as 98.9\%.

In Section \ref{sec:RelatedWork}, an overview of the current state-of-the art and work related to radar-based gesture and sign language recognition is given.  In Section \ref{sec:RFDataAcq}, we describe the RF sensor utilized, experiments conducted, and pre-processing algorithms applied to the data.  In Section \ref{sec:fidelity}, kinematic and replicability considerations are applied to select 15 out of a total of 110 measured ASL signs as example trigger signs.  Next, in Section \ref{sec:motionDetect}, a motion detection method is presented and its efficacy on the acquired datasets demonstrated for temporal segmentation. In Section \ref{sec:multiInputNetwork}, the proposed JD-MIMTL framework for sequential classification is detailed. Results demonstrating the performance of the proposed approach for trigger word detection and sequential recognition are discussed in Section \ref{sec:Results and Discussion}. Discussions of key conclusions and future work is given in Section \ref{sec:Conc}.

\section{Related Work}
\label{sec:RelatedWork}

\subsection{Radar-based Activity and Gesture Recognition}
\label{subsec: Radar-based Activity and Gesture Recognition}

\begin{figure*}[t!]
\centering
\includegraphics[width=17 cm]{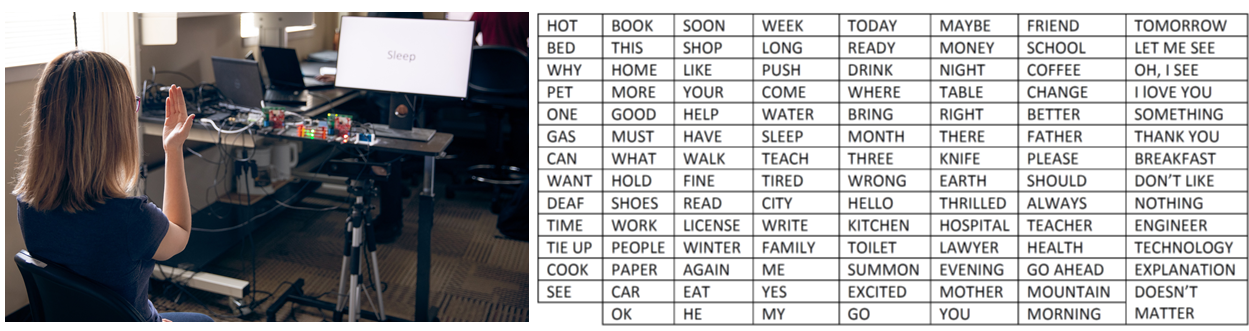}
  \centering \caption{Experimental setup (left) for acquisition of ASL signs (listing on the right).}
  \label{fig:ASL_List}
\end{figure*}

A variety of deep learning approaches \cite{Gurbuz_SPM_2019} have been leveraged for human motion recognition with RF sensors. Most approaches consider either daily activities (e.g. walking, sitting, running) or hand gestures (e.g. left/right, up/down swiping, push buttons).  Recurrent neural networks (RNNs) have been proposed in many works, but results have been primarily demonstrated on fixed-duration snapshots that include just one class of motion.  For example, \cite{Wang_SGRUN2018} applies stacked gated RNNs to 2D micro-Doppler signatures, while  \cite{Latern2018} constructs a 3D data representation from shifted windows of the micro-Doppler signature, applying both a 3D convolutional neural network (CNN) and LSTM.  A more common approach is to use a time series of range-Doppler maps as input \cite{Soli2016,Santra2018} to the 3D CNN-LSTM network, while also using Connectionist Temporal Classification (CTC) \cite{Latern2018} and triplet loss \cite{Santra2019}.   

Studies considering recognition performance in real-world conditions are limited; most of the aforementioned works involve experiments conducted in controlled environments with participants positioned at a fixed distance from the RF sensor.  Work involving real-world use cases have considered the effects of sensor positioning and environment.  For example, \cite{GestureRadar2021}, investigates the dependence of performance on sensor position at different heights from the ground and distance from the user. Dynamic time warping is used on RF data acquired from a dual-Doppler radar to accurately recognize 12 different gestures with at least 80\% accuracy, depending positioning.  These results are consistent with expectations based on the radar range equation, which show that the received power decreases with distance ($R$) as $1/R^{4}$.  Indoor environments also tend to have comparable stationary clutter properties, and, hence do not result in significant variations in performance. The environment-independence of RF sensing of ASL was verified in a recent  \cite{mmASL}, which found the recognition accuracy across several different rooms to be comparable.  Another important factor in gesture recognition is the upper body movements, which can change the received RF signal from gestures.  In \cite{Gu2019}, a single transmitter, dual receiver RF transceiver was proposed to decouple hand gestures from random body movements, and thereby improve gesture recognition accuracy.


An important real-world use case that is gaining increasing attention is the challenge of sequential motion recognition. Human movement is inherently dynamic, greatly varied, and sequential in nature.  The continuous data streams acquired by RF sensors in real-world environments will consist of an intertwining of gross body activities and finer movements, such as gestures. But, most works on human activity recognition consider either daily activities or fine-grain gestures, while the approaches proposed reflect a similar technique to that applied over fixed-duration snapshots.  For example,  \cite{Haobo2021} applies bi-LSTM networks to continuous sequences of micro-Doppler signatures, while \cite{Ding2019} applies RNNs to 
continuous streams of 3D inputs formed from the time series of range-Doppler maps.

\subsection{Radar-based ASL Recognition}
Radar-based ASL recognition to-date has primarily focused on the recognition of snapshots of specific words or phrases. In \cite{Kulhandjian2019}, ten (10) different ASL phrases that would be relevant to emergency response were recognized with an accuracy of 95\% using transfer learning from VGG-16 to classify X-band micro-Doppler signatures. In \cite{Gurbuz2021_ASLR}, feature-level fusion of RF sensors operating at three different transmit frequencies (10 GHz, 24 GHz, and 77 GHz) were used together with a random forest classifier trained only an measured micro-Doppler signatures from fluent ASL signers yielded a classification accuracy of 72.5\% for 20 ASL signs.  Moreover, using a support vector machine (SVM) classifier, it was shown that the ASL articulations of hearing imitation signers were distinguishable from that of fluent ASL users.  With the use of a multi-modal DNN for fusion \cite{Gurbuz_JSEN_2021}, the classification accuracy for 20 signs was improved to 95.5\% and shown to surpass by 22\% and 19\% the accuracy given from use of a single RF sensor classified using transfer learning from VGG-16 and unsupervised pre-training with convolutional autoencoders (CAEs), respectively.  

In \cite{mmASL}, micro-Doppler signatures of 50 ASL signs are classified with an average accuracy of 87\%, while the sign \textsc{KNOCK} is specified as a wake word and detected at a rate of 94\% using a fixed-window binary DNN classifier.  Word-level ASL recognition with RF sensors is shown to be tolerant to the presence of other interfering users, different user positions and different environments. These results were achieved by collecting over 12k samples from 15 different participants.

Because of the differences in fine-grained temporal dynamics and linguistic parameters, such as prosody and grammatical structure, the RF data acquired from hearing imitation signers versus fluent ASL users are actually quite different. In \cite{Gurbuz_JSEN_2021}, it is shown that imitation signing cannot be used to train classifiers of fluent signers.  To overcome this challenge, adversarial learning has been proposed \cite{Mahbub_TAES2021} to 1) adapt imitation signing data to resemble that of fluent signers, and 2) synthesize kinematically accurate samples for training DNNs.  This approach has yielded over 77\% top-1 and over 93\% top-5 accuracy for recognition of 100 ASL signs using micro-Doppler signatures acquired from a 77 GHz RF sensor.

Note that all of the above works classify fixed-duration snapshots of micro-Doppler signatures of ASL.  Thus, this work fills an important gap in current literature by addressing the challenge of trigger sign detection and sequential ASL recognition in continuous RF data streams of mixed motion sequences that are typical of daily living.

\section{RF Data Acquisition and Pre-Processing}
\label{sec:RFDataAcq}
\subsection{RF Sensor}
\label{subsec:RFSensor}
In this study, a TI AWR1642BOOST 77 GHz RF transceiver paired with a DCA1000EVM data capture card were used to record data directly to a laptop.  The TI 77 GHz transceiver is a frequency modulated continuous wave (FMCW) short-range automotive radar that has two transmit (TX) channels and four receive (RX) channels, which offer additional sensing capabilities in comparison to other commercially available RF sensors that may have only 1 TX/RX channel.  The antenna for the sensor has a roughly $\pm$70$^{\circ}$ azimuth and $\pm$15$^{\circ}$ elevation beamwidths.  The sensor was positioned on a small table at a distance of about 1 meter from the ground. 

\subsection{Participants}
\label{subsec:participants}
Although ASL has been used as example motions in some gesture recognition studies \cite{Melgarejo2014,WiFinger2016}, sign language greatly differs from gesturing in that it possesses a much greater degree of physical complexity and Shannon information \cite{malaia2016assessment,borneman2018motion, malaia2020syllable}. Like other complex system-generated signals, raw physical signal from signing data contains information at multiple timescales, spanning phonological, semantic, syntactic, and prosodic cues (\cite{blumenthal2019shared, wilbur2008contributions}). 

While some studies \cite{DeepASL2017,SignFi2018} have utilized imitation signers - i.e., hearing participants who mimic signs observed in video - it has been shown \cite{beal2019hearing} that it takes at least three years before the signing of ASL learners is perceived as fluent by native ASL users.  Imitation signers exhibit greater kinematic variations, erratic cadence and signing errors, especially in replicating repetitive signs.  Indeed, in our previous works \cite{Gurbuz2021_ASLR,9425571}, we have found that imitation signing is distinguishable from native signing using classification of RF $\mu D$ signatures.

Thus, in this work, RF data from both imitation signers and native ASL users were acquired and used for comparative study in trigger sign selection.  A total of 110 single ASL signs were recorded from participants sitting 1 meter away from the radar.  A total of 19 participants contributed to the database, including 4 native ASL users, who were either Deaf or child-of-deaf-adults (CODA), and 6 hearing individuals.  Continuous recordings of mixed activity/signing sequences were recorded from 13 hearing participants, while testing on native users was conducted with 2 CODAs and 2 ASL learners, who were not used in acquisition of training samples. 

\subsection{RF Datasets}
\label{subsec:RFDatasets}
A total of two different datasets were acquired:
\begin{enumerate}
    \item \textbf{Single ASL Signs:}  110 of the more frequently used ASL signs were selected from the ASL-LEX Database \cite{Caselli2017}, including nouns, verbs, and adjectives.  A complete listing of the signs acquired is given in Figure \ref{fig:ASL_List}.  Each participant was asked to repeat the signs 5 times, resulting in 20 native and 30 imitation samples per sign.
    \item \textbf{Mixed Motion Sequences:}  Of these 110 signs, based on kinematics and replicability, a subset of 15 ASL signs are selected (see Section \ref{sec:fidelity}).  Five different sequences of three ASL signs mixed with three different gross motor activities (walking, sitting, and standing up) were acquired, as shown in Table \ref{tab:sequences}.  For example, in \textsc{sequence 1}, the participant first walks for a few seconds, then sits on a chair located in front of the radar and enacts 3 different signs (\textsc{tired, book, sleep}), and finally stands up. The participants were instructed to perform these activities consecutively in the line-of-sight of the radar.  A total of 200 hearing participant samples and 94 native participant samples for each sequence were acquired, and made available for download \footnote{https://github.com/ci4r/ASL-Sequential-Dataset}.
\end{enumerate}

\begin{table}[!t]
    \centering
    \caption{Description of Mixed Activity/Sign Sequences}
    \begin{tabular}{|c|c|}
        \hline
         \textbf{Seq. \#} & \textbf{Motion Sequence} \\
         \hline
         \hline
         1 & Walking, sitting, \textsc{tired, book, sleep}, standing up\\
         \hline
         2 & Walking, sitting, \textsc{evening, ready, hot}, standing up\\
         \hline
         3 & Walking, sitting, \textsc{month, cook, again}, standing up\\
         \hline
         4 & Walking, sitting, \textsc{summon, maybe, night}, standing up\\
         \hline
         5 & Walking, sitting, \textsc{something, teacher, teach}, standing up\\
         \hline
    \end{tabular}
    \label{tab:sequences}
\end{table}

\subsection{Transmit Waveform Parameters}
\label{subsec:transmitWaveformParam}
\begin{figure}[!b]
\centering
\includegraphics[width=8.7cm]{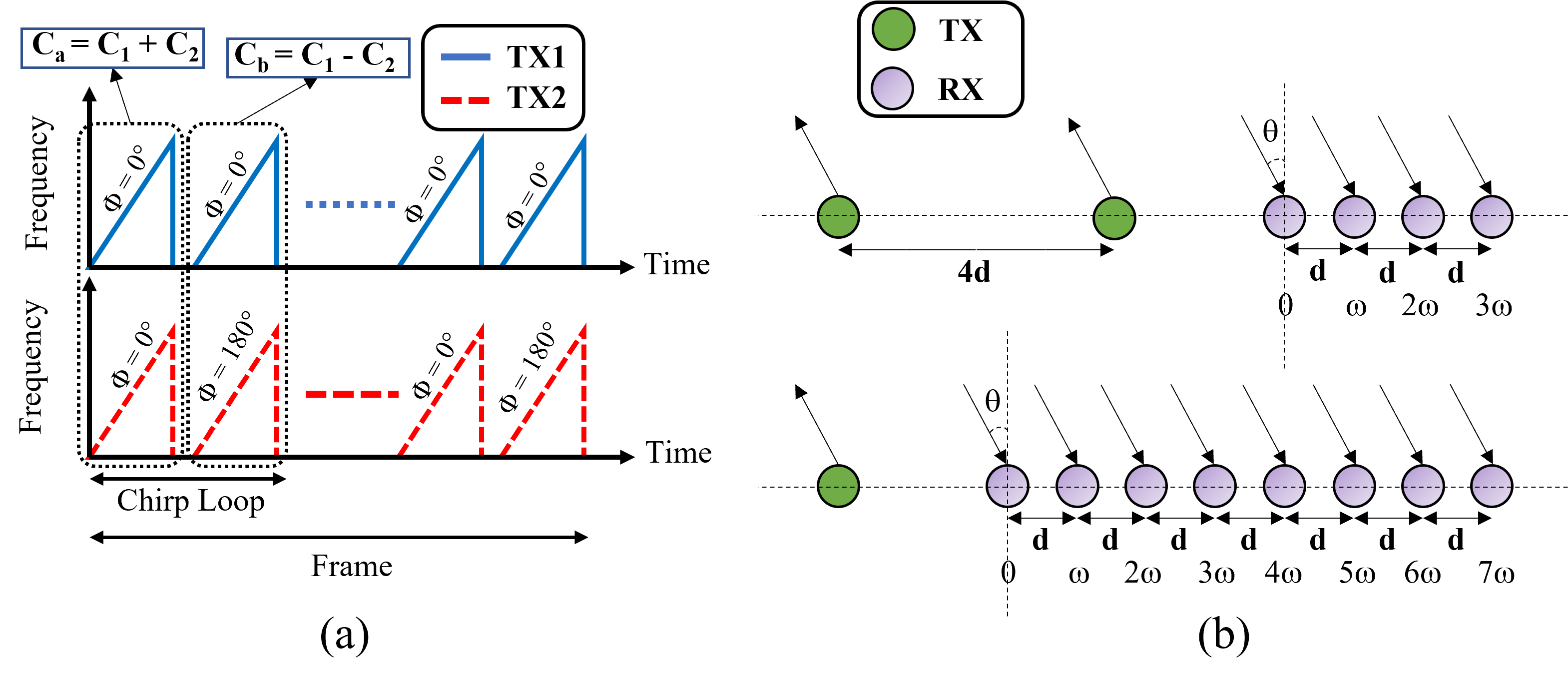}
  \centering \caption{(a) BPM chirp configuration, and (b) virtual array synthesis.}
  \label{fig:mimo}
\end{figure}

The  raw  data  provided  by each receive channel of the  RF  sensor  is  a  time  stream of  complex in-phase (I) and quadrature (Q)  data.  The presence of multiple receive channels enables not only the extraction of range and velocity, but also the direction (or angle) of arrival of the received signals.  The 77 GHz TI transceiver has 2 TX and 4 RX channels, which forms a uniform linear array (ULA).  If, instead, the transmitters send two different chirp combinations, binary phase modulation (BPM) can be used to form a virtual array that behaves like a single TX, but 8 RX channel transceiver.  To accomplish this, in the first chirp,
$C_a$, we send exactly the same chirps from both transmitters, TX1 and TX2, with phase of $\Phi$ = 0\degree, while in the second chirp, $C_b$, TX2 transmits with phase of $\Phi$ = 180\degree. The chirps of TX1 and TX2 then can be retrieved by
$C_1 = (C_a + C_b)/2$ and $C_2 = (C_a - C_b)/2$, respectively.  In this way, we obtain a phase-shift for our second transmitter as well and synthesize a virtual array.  An illustration of the transmitted chirp waveforms are provided in Figure \ref{fig:mimo}a, while the resulting actual and virtual ULAs are shown in Figure \ref{fig:mimo}b. 


For a target at angle $\theta$, the phase difference between receiver channels will be as follows:
\begin{equation}
    \omega = \frac{2\pi dsin(\theta)}{\lambda}
    \Rightarrow \theta = sin^{-1}(\frac{\lambda\omega}{2\pi d})
    \label{eqn:omega}
\end{equation}
The angular resolution, $\theta_{res}$, is given by:
\begin{equation}
    \theta_{res} = \frac{\lambda}{M \times d \times cos(\theta)}
    \label{eqn:angular res}
\end{equation}
where $M$ is the number of channels, so the doubling of M from 4 (real) to 8 (virtual) using BPM improves the angular resolution by a factor of 2.
Thus, the 77 GHz TI transceiver was set to operate in BPM mode with a bandwidth of 4 GHz, pulse repetition frequency (PRF) of 6.4 kHz with 256 samples per pulse and a coherent processing interval (CPI) of 40 ms.

\subsection{RF Data Representations}
\label{sec:RFDataRep}
Typically, the received I/Q data stream from each channel is reshaped into a 3D array, known as the \textit{radar data cube}, with dimensions of fast-time (the number of ADC samples) $\times$ slow-time (the number of pulses) $\times$ channels.  From the radar data cube, several different ways of representing the information acquired by the radar may be formed.  The Fast Fourier Transform (FFT) across fast-time can be used to find the frequency difference, $f_{b}$, between the transmitted and received signals at any instant of time.  If the chirp rate of the frequency modulated waveform is $\gamma$, then the distance, $R$, between the radar and scatterer can be found as 
$R=c f_{b} / 2 \gamma$,
where $c$ is the speed of light.  An FFT across slow-time reveals the velocity of moving scatterers, $v=cf_{d}/2f_{t}$, where $f_{d}$ is the Doppler shift and $f_{t}$ is the transmit frequency. 

Thus, several different 2D data representations may be computed from the radar data cube:
\begin{enumerate}
\item \textbf{Range-Doppler Map:}  The 2D FFT of the slow-time/fast-time data matrix for a single channel can be computed to find a range-Doppler image for each CPI. Because a RD map is computed from all the received returns acquired over a CPI, some researchers have adapted terminology from video processing and refer to the RD map as a \textit{frame} and the CPI as the \textit{frame} duration. Time series of RD maps can be formed to form RD videos. With a CPI comprised of 256 pulses, the resulting video as a frame rate of 25 fps (1/40 ms). 
\item \textbf{Time-Frequency (Micro-Doppler) Map:} While there are many time-frequency transforms that yield the $\mu D$ frequency versus time, the most often used is the spectrogram \cite{9106153}, which is the square modulus of the Short-Time Fourier Transform (STFT) across slow-time. In order to generate $\mu $D spectrograms independent of the subjects' range, cell averaging constant false alarm rate (CA-CFAR) is applied on RD maps for detection of range bins with motion. Detected range bins are then used to generate the spectrograms.
\item \textbf{Range-Angle Map:} 
Angle can be computed from multiple-channel data using a beamforming method, e.g. multiple signal classification (MUSIC), to determine the angle-of-arrival of the received returns at a specific range and Doppler. Repeating this process for each CPI yields a time-series of RA maps, i.e. RA videos.  
\end{enumerate}






\begin{figure*}[t!]
\centering
\includegraphics[width=18 cm]{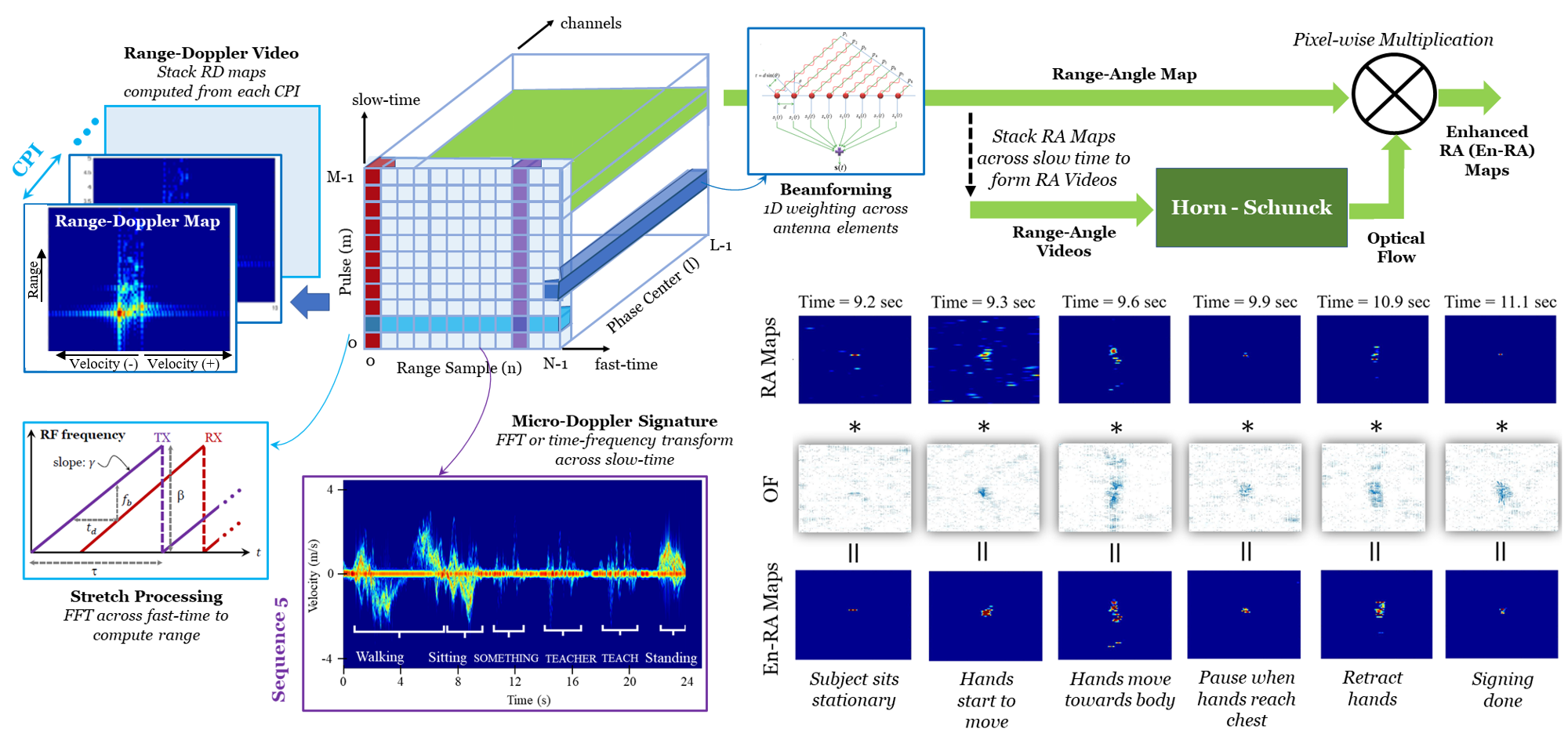}
  \centering \caption{Signal processing diagram for computation of various RF data representations.}
  \label{fig:RFdatareps}
\end{figure*}

The visibility of target-related motion in the RA maps may be enhanced using optical flow, which indicates the spatial change in the location of pixels from one frame to another in a video.  In this work, we compute the optical flow using the Horn-Schunck method \cite{10.5555/888857} and take its element-wise multiplication with the pixels in the RA maps to accentuate motion-related returns.  This process puts more weight on pixels where there is a moving target, and suppresses pixels comprised of clutter or minimal motion.  Because the MUSIC algorithm is relatively prone to noise, this approach can enable significant visual enhancements in the RA maps.  An overview of the radar signal processing steps utilized to compute the stated RF data representations are summarized in Figure \ref{fig:RFdatareps}.

\section{Trigger Sign Fidelity Analysis and Selection}
\label{sec:fidelity}

\begin{figure*}[!t]
\centering
\includegraphics[width=18cm]{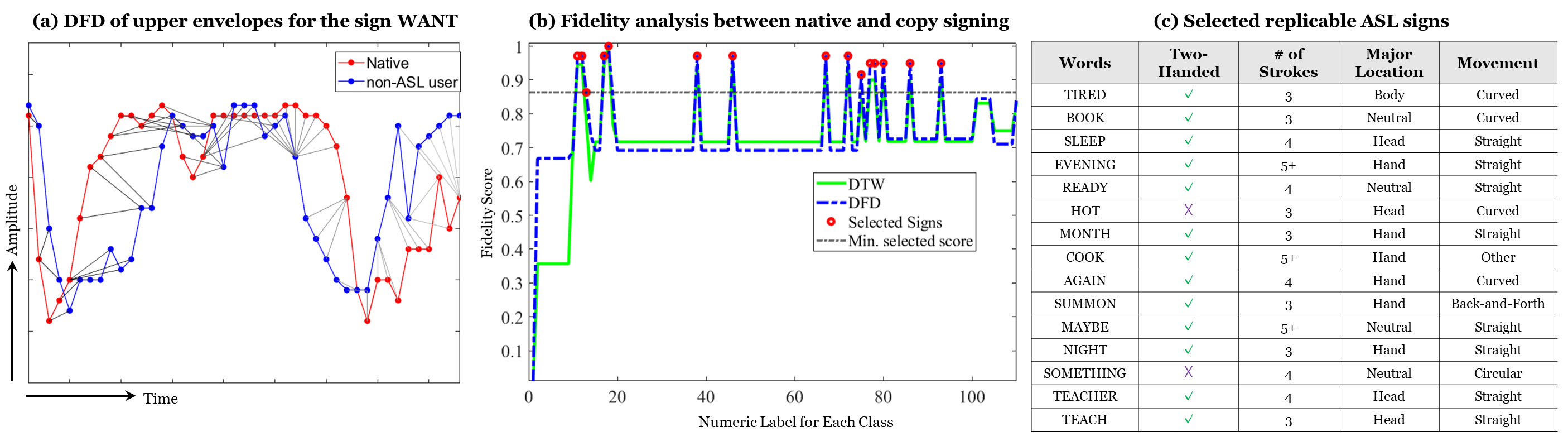}
  \centering \caption{Selection of replicable ASL signs using DFD and DTW.}
  \label{fig:dfd}
\end{figure*}

There are many different considerations for the design of a device trigger sign (also known as a wake word).  Trigger signs should be distinct, not easily confused with signs frequently used in daily discourse, easy to articulate and culturally appropriate.  In Deaf culture, for example, while it is common for finger-spelling to be used to state the names of a hearing individuals, personal \textit{name signs} can only be used if the name sign has been given by a member of the Deaf community.  Moreover, ASL does have some differences in dialects used in different geographical regions within the U.S., such as Black ASL, which represents a unique ethnic sub-culture in the South \cite{Hill2012_BlackASL}.  The cultural context of signs may differ and take on different meanings in different regions.  Therefore, the design of culturally-appropriate trigger signs can only be accomplished through partnership with Deaf community organizations, who can provide cultural perspectives and facilitate studies soliciting Deaf community feedback on the design.

Thus, this paper focuses on technical aspects of trigger sign design as a precursor to a subsequent Deaf-centric design study. First, as RF sensors are sensitive to distance and motion, signs that are dynamic, with strong radial velocity components (i.e. include primary arm motion, as well as secondary motion of the hand, such as handshape or orientation change), or which traverse greater distance and have a longer flight times are better suited as trigger signs for automatic detection.  This is in contrast with signs primarily characterized by secondary hand motion, such as fingerspelled words.  

Second, the replicability of the trigger sign is important to enable consistent and robust recognition.  Although native ASL users are the target population for ASL-sensitive user interfaces, there is a wider community of ASL learners and non-native ASL users, such as interpreters, who could also be using the interface.  However, as noted in Section \ref{subsec:participants}, there can be noticeable differences in the articulation of signs based on fluency.  Thus, the replicability of the 110 signs listed in Figure \ref{fig:ASL_List} were evaluated using a comparison of the imitation signing and native ASL $\mu D$ signatures.  This was done by first computing the upper and lower envelopes of each sign based on 
the percentiles of the cumulative amplitude distribution \cite{4469865,7165625}.  Next, both the Discrete Fr\'echet Distance (DFD) \cite{dfd} and Dynamic Time Warping (DTW) were used to compare the replicability of signs based on fluency.  

DTW is a method for measuring the similarity between two time-series and finds the optimal match \cite{optmatch} between sequences that satisfy all restrictions and rules with the minimum cost.
The DFD computes the similarity between two curves by taking into account both ordering of the points and the location along the curves. It is defined as the shortest cord-length required to join a point traveling forward along one curve and one traveling forward along the other curve, and the rate of travel for either point may not necessarily be uniform. As the similarity of two curves increases, DFD gets closer to zero. As an example, consider the comparison of the upper envelopes of the $\mu D$ signatures for imitation signing and native signing   for the sign \textsc{want}, shown in Figure \ref{fig:dfd}(a), where the grey lines represent the cord-length.  

To identify the most easily replicable signs (independent of fluency), the envelopes of the native ASL signatures and those from hearing imitation signers are compared on a sign-by-sign basis.  The DTW and DFD metrics are averaged and re-scaled between 0 and 1.  Once the distance metrics, $dtw$ and $dfd$, are normalized, the fidelity scores, $s_{dtw}$ and $s_{dfd}$, for each class (sign) are found by taking the inverse of the normalized distance (i.e., $s_{dtw} = 1/dtw$, $s_{dfd} = 1/dfd$).  The results are shown in 
 Figure \ref{fig:dfd}(b). It may be observed that both the DTW and DFD are consistent in their assessment of which signs are consistently articulated across deaf, CODA, and hearing users.  
 
 The top 15 signs that have the shortest distance (i.e. highest similarity) between native ASL and imitation signing users
were selected as trigger sign candidates, which will next be evaluated based on detection rate and sequential recognition accuracy.  The selected signs are listed in Figure \ref{fig:dfd}(c) along with their kinematic properties, as given by ASL-LEX.

\section{Motion Detection and Segmentation}
\label{sec:motionDetect}

\begin{figure*}[!t]
    \centering
    \begin{subfigure}[t]{0.45\textwidth}
    \includegraphics[width=1\linewidth]{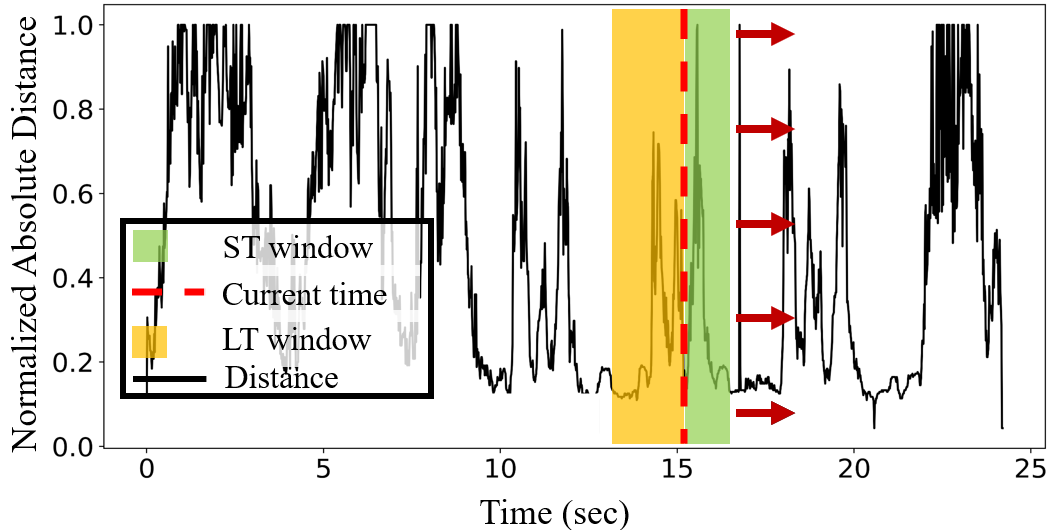}
    \caption{Operation of detector on absolute distance vectors.}
    \label{subfig:stalta detect}
    \end{subfigure}
    \centering
    \begin{subfigure}[t]{0.45\textwidth}
    \includegraphics[width=1\linewidth]{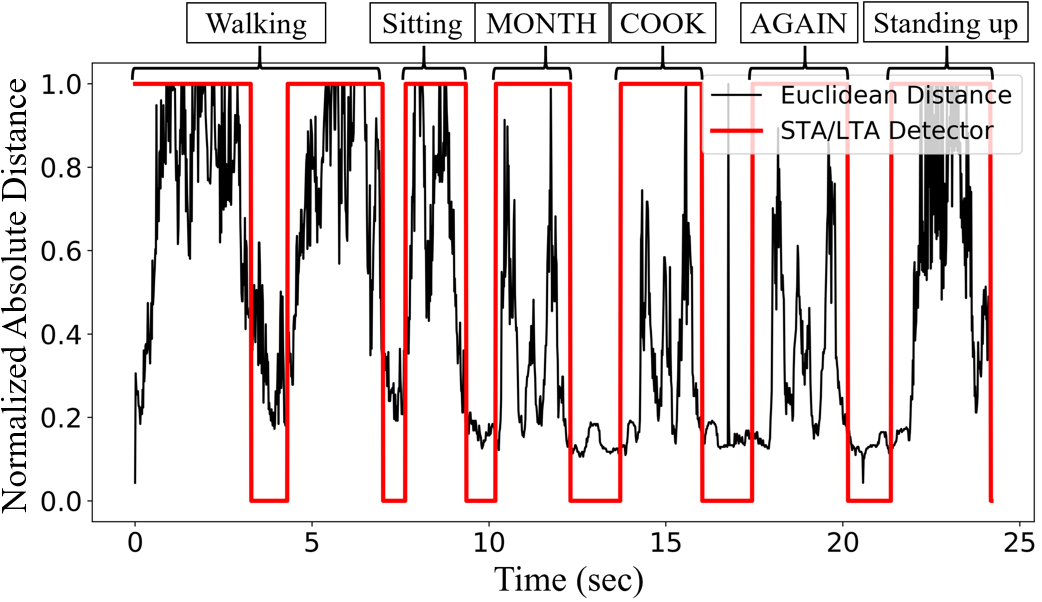}
    \caption{Intervals with motion detected by STA/LTA detector.}
    \label{subfig:stalta detected}
    \end{subfigure}
\caption{Illustration of the operation of STA/LTA based motion detector on \textsc{Sequence 3}.}
\label{fig:stalta general}
\end{figure*}

Continuous activities and ASL signing create a time series of sequential activities, for which segmentation is an important initial step in the analysis of sequential data.  Utilization of a motion detector can facilitate segmentation, which helps define the length of the input samples to be fed to a learning model.  It can also improve the power and computational efficiency of the system by making a prediction only when an activity or sign is detected as opposed to every time step.  While motion detection can be done with a human-in-the-loop approach, this is not desirable in automate, stand-alone systems.  Instead, a power-based automated segmentation algorithm, such as \textit{short time average over long time average} (STA/LTA) \cite{8188276,Sun2020}, dynamic boundary detection (DBD) \cite{9531622} or power burst curve \cite{rs12030454} (PBC) may be utilized. 

The PBC can be used for motion detection using thresholding. The start and end of the motion is determined by when the input power exceeds or falls below this threshold, respectively. An important drawback of this method, however, is that it is prone to a high rate of false triggering, especially in the presence of noise, because the threshold is not adaptive and unaware of past and future power levels. 

STA/LTA-based techniques solve this problem by defining two consecutive windows; namely, short-time and long-time windows. Their relative average power is used to define an adaptive threshold value. The STA/LTA method proposed in \cite{Sun2020} has proven to be very successful in detecting the tail (end point) of hand gestures. However, the method uses fixed length detection windows, whose duration is selected based on the duration of the longest gesture in the dataset. This approach is not well suited to sign language, since ASL signs possess great variability in duration. Basing window size on the longest duration sign can result in a long blank period at the beginning of the detected region for short signs, thereby introducing non-informative or redundant input to the feature space.

DBD, on the other hand, requires application of high-pass filtering to the Doppler information, resulting in elimination of the low and zero frequency components of the spectrograms. Prior work \cite{Gurbuz2021_ASLR} has shown, however, that filtering at 77 GHz results in significant loss of low-frequency information in the signal, together with removal of the clutter, thereby degrading classification accuracy.

Thus, this work proposes a variable window STA/LTA-based motion detection algorithm to identify both the starting and ending point of a motion. First, the absolute difference between the upper and lower envelopes at a time index is computed to create absolute distance vectors. An exemplary, normalized absolute distance vector is shown in Figure \ref{subfig:stalta detect}. The absolute distance for each data recording, $i$, can be computed as $ v_i = |u_i - l_i|$, where $v_i$ is the absolute distance vector, $u_i$ and $l_i$ are the upper and lower envelopes, respectively.

Then, $STA(t)$ and $LTA(t)$ can be defined as the leading and lagging windows at time $t$ as:
\begin{equation}
    STA(t) = \frac{1}{T_1}\sum_{k = t+1}^{t+T_1}{v_i(k)}, 
    \hspace{0.1cm}
    LTA(t) = \frac{1}{T_2}\sum_{k = t-T_2+1}^{t}{v_i(k)} 
\end{equation}
where $T_1$ and $T_2$ are the lengths of short and long windows respectively.  The starting point of a motion is detected when the following conditions are satisfied:
\begin{equation}
    STA(t) > \sigma_1 \hspace{0.3cm} \textrm{and} 
    \hspace{0.3cm}
    \frac{STA(t)}{LTA(t)} > \sigma_2
    \label{eqn3}
\end{equation}
where $\sigma_1$ and $\sigma_2$ are predefined detection thresholds. Similarly, the ending point is detected if 
\begin{equation}
    STA(t) < \sigma_3 \hspace{0.3cm} \textrm{and} 
    \hspace{0.3cm}
    \frac{STA(t)}{LTA(t)} < \sigma_2
    \label{eqn4}
\end{equation}
where $\sigma_3$ is the detection threshold for the stopping point.

Note that in order to locate the starting point, according to \eqref{eqn3}, $STA(t)$ needs to exceed the threshold $\sigma_1$, implying that the the motion has to appear in the short window. Also,
the ratio of average power in the short and the long window should be higher than $\sigma_2$.  In this way, if there is noise, the system will not be triggered unless the ratio exceeds the $\sigma_2$.  Similar conditions apply to ensure correct detection of the endpoint; i.e., the case when the motion disappears from the proceeding window and the ratio drops below the threshold $\sigma_2$. The resulting detection mask found with the proposed \textit{vw}-STA/LTA approach is able to separate the intervals with and without motion, as shown in Figure \ref{subfig:stalta detected}.

\begin{figure}[!b]
\centering
\includegraphics[width=9cm]{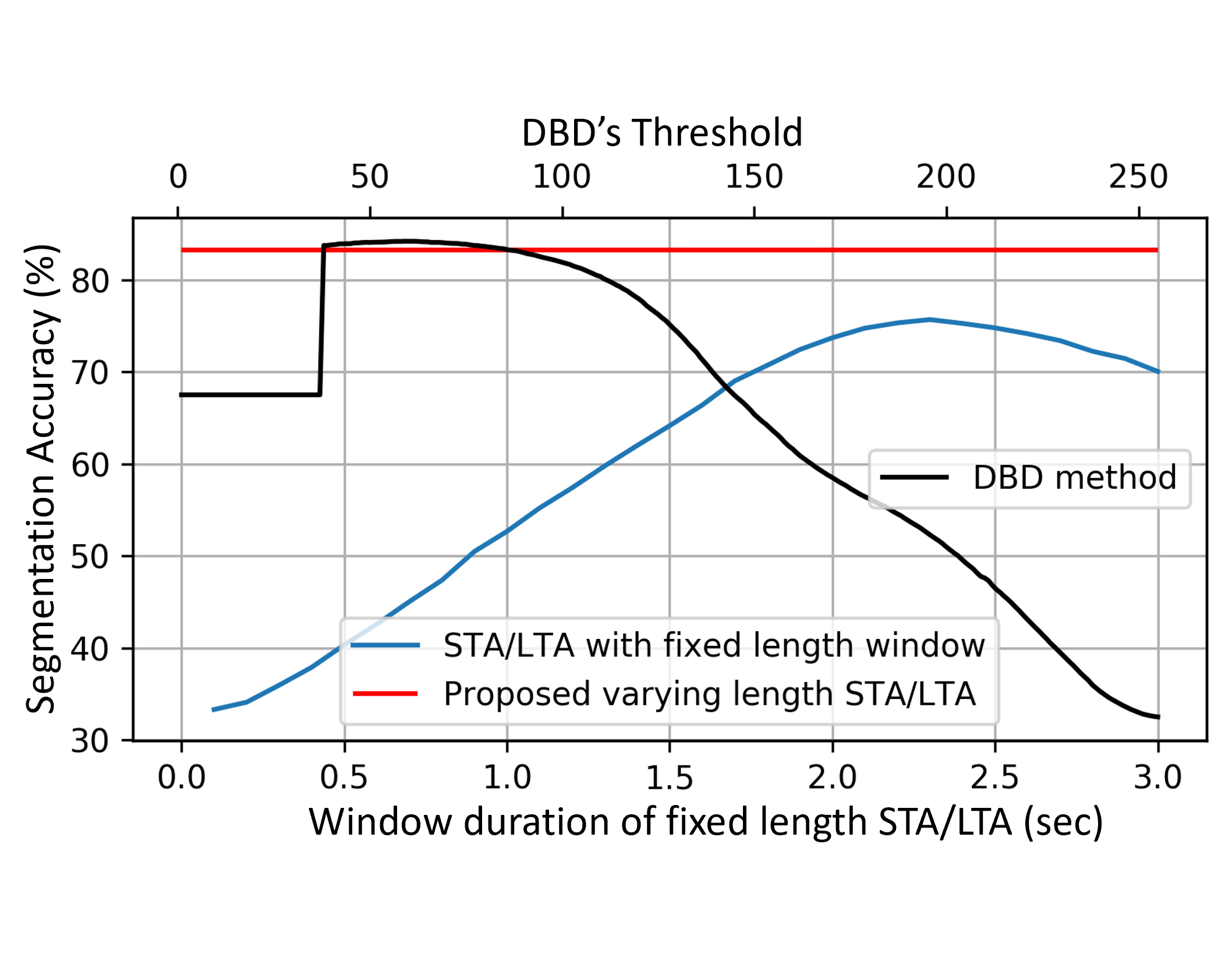}
  \centering \caption{Comparison of the segmentation accuracy of DBD, fixed-window STA/LTA and the proposed variable-window STA/LTA.}
  \label{fig:3 detectors}
\end{figure}

While DBD requires the optimal selection of a threshold based on the returned signal strength, fixed length STA/LTA bases selection on the window length. In contrast, the proposed variable length STA/LTA approach adaptively changes its detection window interval irrespective of the returned signal strength. A comparison of the segmentation accuracy for these three methods is presented in Figure \ref{fig:3 detectors}. Segmentation accuracy is computed by comparing segmentation mask with the ground truth generated by a human analyst for each time step. Note that the segmentation accuracy of DBD and fixed-window STA/LTA exhibit great variance in efficacy for different thresholds or window lengths.  Fixed-window STA/LTA achieves a peak accuracy of 75.7\% when the window length is 2.3 seconds. DBD performs better by comparison, achieving a peak accuracy of 84.2\% when the threshold is set to 61, but with the cost of information loss in low frequency components (see Section \ref{subsec: motion detectors}). This peak value is only slightly higher than the 83.5\% accuracy achieved by the proposed motion detector, while the propose approach can maintain this accuracy irrespective of any parameter values due to the use of variable, adaptive window lengths.

\section{Multi-Input Multi-Task Learning Framework}
\label{sec:multiInputNetwork}

Conventional approaches to RF signal classification rely on a single data representation, presented as either 2D or 3D inputs.  In contrast, to take advantage of all available physics-based information (range, velocity, frequency and angle),  we propose a JD-MIMTL-based DNN architecture, where each input representation is processed in parallel and the final feature space is constructed by fusing individual feature spaces. Auxiliary tasks are used to regularize and better guide the training loss. The accuracy of the proposed approach surpasses that of conventional single-input models by over 13\%.

\subsection{Mixed Motion Sequential Recognition}
\label{subsec:mixedMotionSeqRecog}
Sequential classification of daily activities and ASL signs differs from conventional hand gesture recognition tasks because it is not comprised of just an isolated, short duration, single type of motion. Instead, it consists of a time series of consecutive motions, which might belong to different classes of gross daily activities or ASL signs. A typical approach to classify a continuous time series data includes: 1) temporal segmentation, 2) making prediction for each time step. The former is achieved using a motion detector described in Section \ref{sec:motionDetect}, while the latter will be discussed in this section. In real-world scenarios, training a model with the entire stream of data sequences (24 sec each) is not feasible, because this significantly increases the computation time, rendering outputs only after a long delay, which is undesirable in interactive systems. 
However, when models are trained with shorter input sequences, performance also tends to drop gradually, because performance of LSTMs are dependent on input sequence lengths \cite{seqlen}. Since LSTM networks have the flexibility to  be trained with varying sequence lengths, the data segments isolated by the motion detector were used as input sequences.  These segments will have varying lengths depending on the user's pace and the motion itself.

\subsection{Training a spatio-temporal model}
\label{subsec:trainingSpatioTemp}

In this section, the effect of input sequence length on prediction accuracy is examined. For this purpose, we use a DNN consisting of 3 time-distributed (TD) 2-D convolutional blocks with kernel sizes of 3, followed by max pooling layers and a bidirectional long-short-term-memory (BiLSTM) layer. A TD \textit{softmax} layer is employed for temporal classification. While convolutional layers extract the spatial features, the TD wrapper enables application of the same nested layer to each time step. BiLSTM is a kind of recurrent neural network which is used to extract temporal relationships between time steps. They have proven to be very successful in terms of learning long term dependencies in various tasks such as natural language processing \cite{DBLP:journals/corr/abs-1708-02709}, and speech recognition \cite{6638947}. By employing LSTMs in our final encoded feature space, both spatial and temporal features are extracted for classification.

In $\mu$D spectrogram ($\mu$DS) classification, spectrograms are divided into 0.2 sec \textit{non-overlapping} windows to be used as time steps.  In RD and RA map classification, the interval between each RD/RA map or frame is 40 milliseconds, so to obtain a data structure corresponding to the same (0.2s) duration, five RD/RA frames were stacked (5$\times$40ms = 0.2s). For both inputs, 80\% of the data is used for training and 20\% for testing, with an equal number of samples from each sequence. Adam optimizer and categorical cross entropy is used along with early stopping with patience of 10 epochs to train the model. 
 Hence, the input data has the shape of \texttt{(batch size, number of windows, width, height, channels)}. 
A 2D-CNN+BiLSTM network for $\mu$DS and 3D-CNN+BiLSTM network for RD/RA maps are employed.  The impact of the motion detector is discussed next.

\subsubsection{Original Sequential Data}  
\label{subsec:no motion detect}
Table \ref{tab:red no det} shows the classification accuracy for each input data representation as a function of various input durations. It may be observed that the accuracy of the models for all input domains decreases as the length of input sequences gets shorter. Best performances are obtained using longest sequences with RD maps providing a 92.4\% accuracy. The performance using $\mu$DS changes around 17\% while that using RD maps and RA maps change around 20\% from 1 sec. sequences to 24 sec. sequences. While the longer sequences give better performance, they also result in greater prediction delay and higher memory requirement due to increased data size. This situation demonstrates the challenge of deciding an appropriate input length while doing sequential classification and the trade-off between prediction performance and delay. 

\subsubsection{Motion Detected Intervals (MDI)} 
The detector extracts data segments containing motion, eliminating periods of no movement. Thus, each MDI is of varying duration, and models are trained using variable length data.  
The testing accuracies obtained when using $\mu$DS, RD and RA maps are 78.8\%, 72.8\%, 67.5\% respectively. These results are comparable to those obtained with fixed length sequences of 2 sec. for $\mu$D, and 1 sec. for RD/RA maps, while the length of detected segments vary between 0.6 and 10 sec. Moreover, using MDI rather than fixed length windows significantly reduces the computation time for prediction by masking out the intervals that do not contain any motion. Table \ref{tab:comp time} presents the total computation time of an NVIDIA Titan V GPU to make predictions for data durations of 1 sec and 2 sec. The total computation time is reduced by 45\% on average for different input representations when compared with 2 sec length sequences. Note that the amount of computational savings obtained using the motion detector does depends on the data, in that as MDI increases so does the time savings.  As daily life often involves extended stationary periods, in practical settings the use of MDI can result in significant savings.

\begin{table}[!t]
\centering
\caption{Sequential Classification with CNN+BiLSTM}
\begin{tabular}{ |c|c||c|c|c| } 
  \hline
   Data &  \makecell{Length of \\ Sequences}
   & $\mu$D Spectrogram  & RD Map  & RA Map \\ 
  \hline
 \hline
 \parbox[t]{5mm}{\multirow{6}{*}{\rotatebox[origin=c]{90}{\makecell{Original \\ Sequences}}}} & 1/24 (1 sec) &  69.2\% & 72.5\% & 69.9\% \\ 
 \cline{2-5}
  & 1/12 (2 sec) &  78.6\% & 76.3\% & 73.7\% \\
 \cline{2-5}
 & 1/6 (4 sec) &  81.3\% & 82.4\% & 79\% \\
 \cline{2-5}
 & 1/3 (8 sec) &  84.3\% & 89.9\% & 85.9\% \\
 \cline{2-5}
 & Half (12 sec) &  84.6\% & 90\% & 87\% \\
 \cline{2-5}
 & Full (24 sec) &  86.1\% & 92.4\% & 89.7\% \\
 \hline
 MDI & Varying & 78.8\% & 72.8\% & 67.5\% \\
 \hline
 
\end{tabular}
\label{tab:red no det}
\end{table}

\begin{table}[!t]
    \centering
    \caption{Computation Times Spent for Prediction}
    \label{tab:comp time}
    \begin{tabular}{c||c|c|c}
        Length of Sequences & $\mu$D Spectrograms & RD Map & RA Map \\
    \hline
    1 second & 201.8 sec & 207.5 sec & 205.8 sec\\
    \hline
    2 seconds & 111.7 sec & 125.7 sec & 123.1 sec \\
    \hline
    Detected Intervals  & 61.4 sec & 69.3 & 67.8\\
    \end{tabular}
\end{table}

\subsection{Effect of Motion Detector on Classification Accuracy}
\label{subsec: motion detectors}

The performance of DNN models rely heavily on the data presented at the input, which in turn is extracted based upon the starting and ending points of the MDIs as determined by the motion detector.  Thus, the ability of a motion detector to accurately extract intervals containing movement impacts the efficacy of classifiers.  Table \ref{tab:detectors accuracy} compares the classification accuracy attained from different input representations extracted using DBD, fixed-length STA/LTA and the proposed variable-length STA/LTA motion detectors.
It may be observed that the proposed variable-length STA/LTA detector yields greater classification accuracy in comparison to other approaches, surpassing fixed-length STA/LTA by 0.4-2\% and DBD by 1.3-6.4\%.  Note that the relatively worse accuracy of DBD is due to information loss incurred during the high-pass filtering, which removes low-frequency signal as well as clutter components, and hence degrades the resulting classification accuracy.



\begin{table}[!t]
    \centering
    \caption{Classification Accuracy of the Motion Detectors}
    \begin{tabular}{|c|c|c|c|}
        \hline
         Motion Detector & $\mu$D Spectrogram & RD Map & RA Map \\
         \hline
         DBD & 72.4\% & 70.9\% & 63.8\% \\
         \hline
         Fixed STA/LTA & 76.8\% & 71.5\% & 67.1\% \\
         \hline
         Varying STA/LTA & \textbf{78.8\%} & \textbf{72.8\%} & \textbf{67.5\%} \\
         \hline
    \end{tabular}
    \label{tab:detectors accuracy}
\end{table}

\subsection{Proposed Approach: JD-MIMTL}
\label{subsec:multi-input-CTC}

To improve the classification accuracy obtained with just one input representation, this paper proposes utilizing fusion of multiple input representations in a multiple-task learning \cite{multitask} framework with connectionist temporal classification (CTC) \cite{Graves06connectionisttemporal}. Although MTL has been implemented successfully in  computer vision \cite{7410526} and natural language processing \cite{10.1145/1390156.1390177}, these applications all involve a single data representation (image, text, speech signal).  In RF sensing, the various physical variables measurable by radar - namely, range, $\mu $D, and angle versus time - are reflected in different data representations, to base recognition decisions on all physical properties, multiple inputs to MTL are advantageous.  The joint feature space derived from multiple input representations is enriched by fusing in a concatenation layer.

MTL jointly optimizes multiple objectives by exploiting domain-specific information contained in commonalities and differences across tasks. By sharing representations among related (auxiliary) tasks, the generalization capability of the model can be improved on the main task.  ASL classification can be aided by basing decisions on consistency with certain physical properties of signing, based on the categorization provided in Figure \ref{fig:dfd}(c).  Five auxiliary tasks are defined: 
\begin{itemize}
\item Task 1: One versus two handedness;
\item Task 2: Major location of hands;
\item Task 3: Movement type;
\item Task 4: Daily activity versus ASL sign; and
\item Task 5: Number of strokes.
\end{itemize} 

\begin{figure}[!t]
\centering
\includegraphics[width=7.9cm]{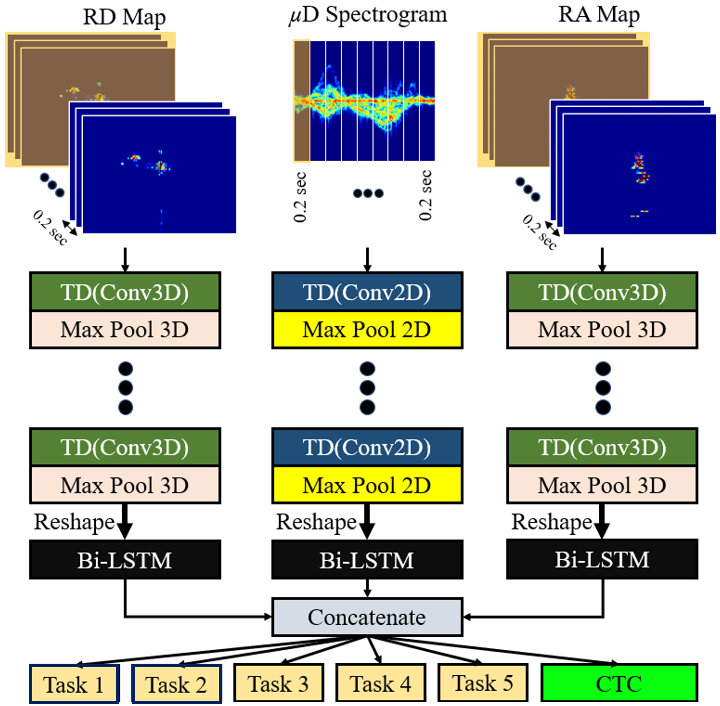}
  \centering \caption{Proposed multi-input multi-task learning network.}
  \label{fig:network}
\end{figure}

The overall loss function, $L_{total}$, utilized in the JD-MIMTL framework is the weighted sum of the CTC loss, $\lambda_{ctc}$, and the loss $L_i$ specific to each task $i$:
\begin{equation}
    L_{total} = \lambda_{ctc}L_{ctc} +  \sum_i^{I}\lambda_i L_i    
\end{equation}
where $\lambda$ are the weights assigned to the various loss terms.
Since each task has its own loss function, and, hence, varying convergence times, the weights  $\lambda$ needs to be jointly optimized. Three different loss optimization techniques \cite{8848395} were compared, namely, the uniform combination of losses (i.e. equal weights across all tasks), the uncertainty based weighing method \cite{8578879}, and grid search. The first two methods minimize $L_{total}$ without taking into account the importance of each individual task.  Since we aim to  minimize $L_{ctc}$, which is derived from the prediction layer, the grid search method was preferred.  The use of smaller auxiliary task weight values during grid search was found to perform better than that obtained with using the uniform combination of losses or uncertainty-based weighting. Specifically, weight values of $\lambda_{ctc}$ = 1 and $\lambda_{i}$ = 0.2 were used.  The overall proposed JD-MIMTL approach is depicted in 
Figure \ref{fig:network}. After training the model, all of the auxiliary task and CTC output layers are removed and the model is augmented with a $softmax$ layer for classification.

The probability distribution of the classes, which is obtained as the output of the JD-MIMTL, can be decoded two ways in parallel for sequential classification and trigger word detection. Best path decoding is used as the decoding scheme of the CTC outputs for both objectives. However, the final prediction class is defined as the statistical mode of the time steps of an MDI for sequential classification, and as the prediction scores for the trigger sign accumulated over the time steps of an MDI for trigger word detection.

\section{Results and Discussion}
\label{sec:Results and Discussion}

\subsection{Trigger Word Detection}
\label{sec:triggerWordDet}

\begin{figure*}[!t]
    
    \centering
    \begin{subfigure}[t]{.3\textwidth}
    \includegraphics[width=1\linewidth]{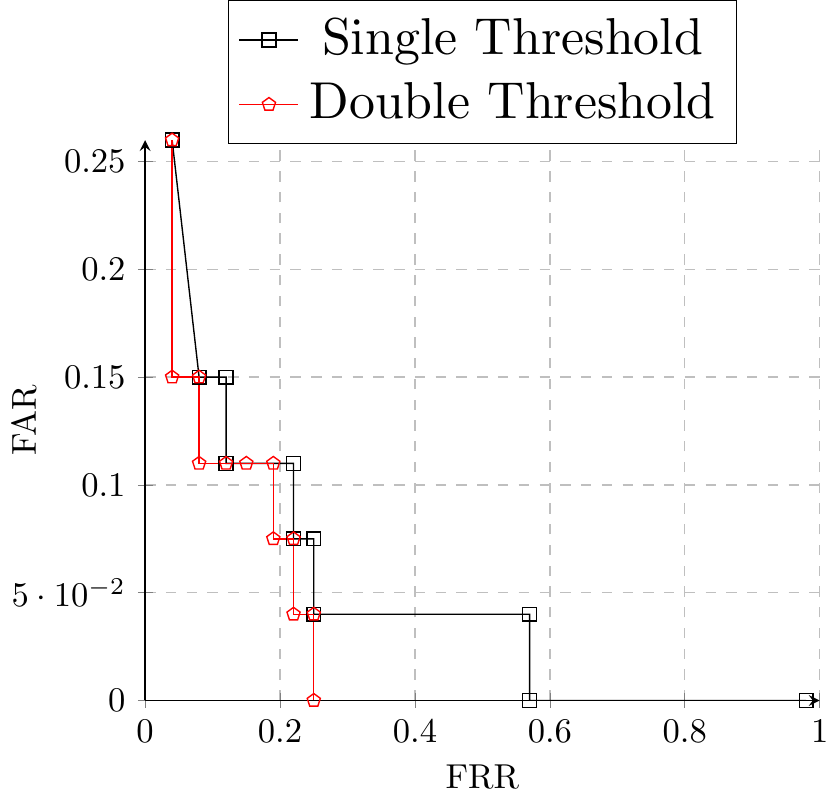}
    \caption{FAR and FRR of the word \textsc{again} for single and double threshold.}
    \label{fig:again th}
    \end{subfigure}
    \centering
    \begin{subfigure}[t]{0.3\textwidth}
    \includegraphics[width=1\linewidth]{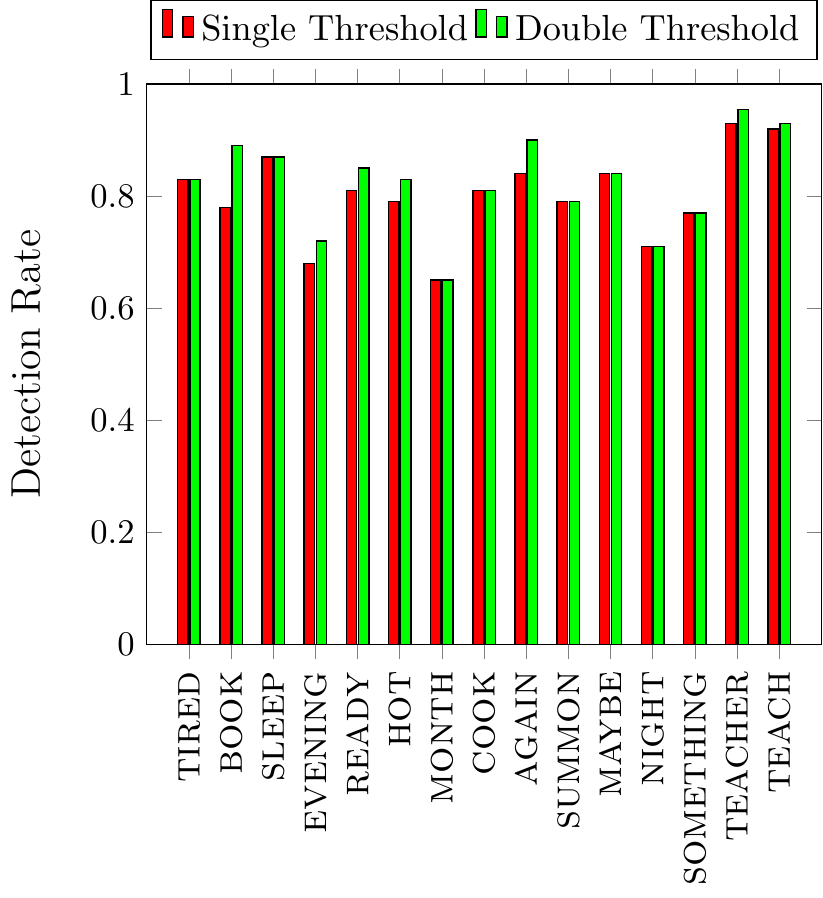}
    \caption{Detection rates of the words for single and double threshold.}
    \label{fig:detection rates}
    \end{subfigure}
    \centering
    \begin{subfigure}[t]{0.3\textwidth}
    \includegraphics[width=1\linewidth]{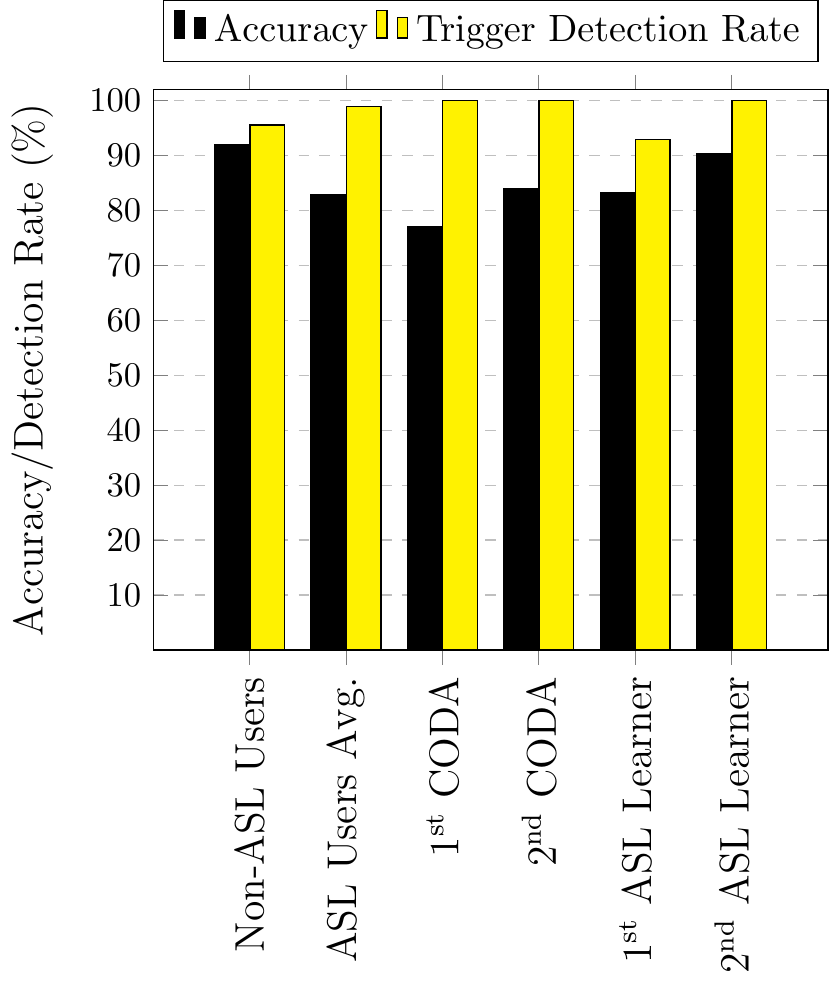}
    \caption{Performance of the proposed method on different ASL fluency groups.}
    \label{fig:native test}
    \end{subfigure}
\label{trigger_general}
\caption{Trigger word detection results.}
\end{figure*}

To activate a device, the trigger sign must be correctly recognized from within a stream of data, and the activation should occur when the articulation of the sign is completed.
One approach is cumulative score aggregation (CSA) \cite{siri}, where the scores (i.e., prediction probabilities) of the trigger sign are accumulated over time, and a detection is recorded when the accumulated score, $s_a$, exceeds a predefined threshold. The threshold can be adjusted to ensure the detection is triggered only when the trigger sign is complete.

In this work, an adaptive, double-threshold CSA approach is proposed for trigger sign detection.
Since the MDIs have varying lengths, the value of the threshold, $T$, is adaptively determined based on the interval length as: $T = w*\gamma$,
where $w$ is the length of the MDI and $\gamma$ is a predefined confidence factor. To mitigate the false rejection rate (FRR) of the detector, a second (lower) threshold, $T_{low}$, is also defined. When the accumulated score exceeds the $T_{low}$, but not $T$, the detector is alerted to the possibility of a trigger and begins recording the duration over which the score stays above $T_{low}$.  The system is triggered if score exceeds $T_{low}$ for more than $w/2$ seconds and the motion is classified as the trigger sign.    



In trigger word detection, effect of using single versus double thresholding can been seen from
Figure \ref{fig:again th}, which shows the trade-off between the false alarm rate (FAR) and FRR for  $\gamma \in \{0.01:0.99\}$ for the word \textsc{again}.  When a single threshold is used, the FRR can climb as high 0.6, while double thresholding limits this value to just over 0.2.  This is significant because decreasing the FRR boosts the detection rate, $D_r = 1 - FRR - FAR$,
where FRR and FAR are defined as:
\begin{equation}
    FRR = \frac{n_{t}-n_{d}}{n_{t}}, \hspace{0.7cm} FAR = \frac{n_{f}}{n_{t}}
\end{equation}
where $n_t$, $n_d$ and $n_f$ are the number of total, detected and false detected samples respectively.

As shown in Figure \ref{fig:detection rates}, when the resulting detection rates for single thresholding versus the proposed double thresholding approach are compared, it may be observed that for each considered trigger sign, the proposed approach yields a same or improved detection rate.
The word \textsc{teacher} has the highest detection rate for both thresholding methods, achieving a detection rate of 0.93 and 0.96, while the word \textsc{month} (self-occluded) has the lowest score of 0.65 for both cases. Signs with higher classification accuracy tend to have higher detection rates as well, such as \textsc{teacher} and \textsc{teach}.  

The number of strokes (i.e. length) of the sign is an important consideration in trigger sign selection. For the purposes of automatic detection, strokes were defined as components surrounding the sign-initial and sign-final handshapes; thus, both the motion inherent to the sign (i.e. the \textit{stroke} as defined in sign language phonology), and transitional motions preceding and following the sign, were included in the analysis. This approach approximated predictive processing in human sign language recognition (\cite{malaia2017current, ford2021classification}), while remaining consistent with ecological paradigm of wake sign use.  Signs with few strokes defined in this manner (less than 3) were found to have many false alarms, while those with more than 4 were prone to a high number of false rejections.  This is similar to results in speech recognition, which report optimal wake word lengths of 3 to 4 syllables \cite{9027708} - or, in quantitative terms, several entropy (high information-density) peaks within the continuous signal.

\subsection{Sequential ASL Recognition}

A testing accuracy of \textbf{92\%} is achieved using the proposed JD-MIMTL approach, and surpasses the results achieved with various state-of-the-art sequential recognition approaches, as shown in Table \ref{tab:seq class table}.  This result is also quite close to the 93.5\% accuracy attained using JD-MIMTL when the motion detector is replaced with ground truth segmentation. Moreover, the baseline established in Section \ref{subsec:trainingSpatioTemp} using CNN+BiLSTM on single-input representation MDI data is improved to 84.3\% by application of feature-level fusion.  Consideration of CTC loss improves the results obtained for both single-input and fusion of multi-input representations.   
The accuracy using $\mu$DS increased to 80.6\%, RD maps to 78.4\% and RA maps to 71.3\%, thus providing an average improvement of 3.73\%.  For RD maps and RA maps, MTL only slight improves performance by just 0.1\%-0.2\%, while the accuracy with $\mu$DS increases by 3\%.  The proposed JD-MIMTL approach yields a performance improvement of 8.4\% over $\mu$DS as a single-input to MTL, and 4.5\% improvement over multi-input feature level fusion without using MTL. 



\begin{figure}[!t]
\centering
\includegraphics[width=8.7cm]{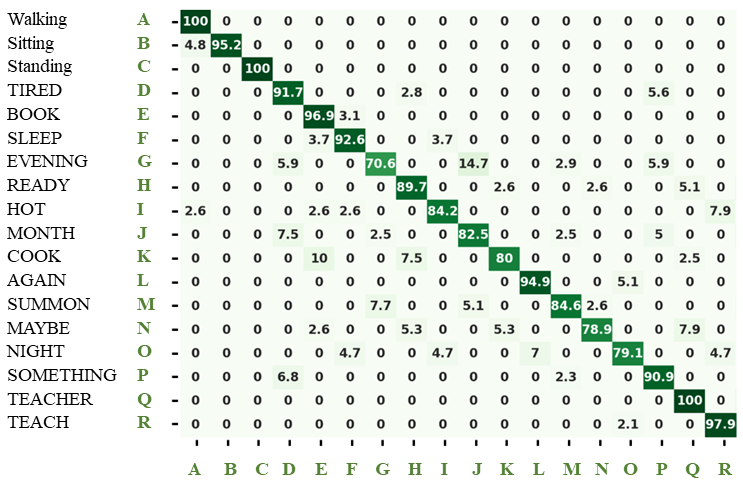}
  \centering \caption{Confusion matrix of the proposed JD-MIMTL.}
  \label{fig:conf}
\end{figure}

  The confusion matrix for the proposed architecture is provided in Figure \ref{fig:conf}. It can be seen JD-MIMTL exhibits the most confusion in signs with low radial motion (\textsc{evening}, \textsc{maybe}, \textsc{night}) and self-occlusion (\textsc{month}). The signs with high radial motion (\textsc{teacher}, \textsc{teach}) have the highest recognition rates. This is due to higher sensitivity of radars to radial velocity components. 







\subsection{Performance Across Different Fluency Groups}
\label{subsec:robustness}
The proposed approach is tested on different fluency groups to evaluate is efficacy across different users.  This is done by training the model solely with data from non-ASL users, but testing on ASL users' data.  Thus, not only are the participants between training and test sets different, but also their fluency levels. In Figure \ref{fig:native test}, the overall testing accuracy for all signs, and the trigger detection rate for the selected trigger word, \textsc{teacher}, are presented for different fluency groups.  While the first two columns report average results, the remaining 4 columns break down the results for specific participants, indicating whether the participant was an ASL learner or CODA. On average, the sequential ASL classification accuracy for ASL users was 10\% less than that attained from non-ASL users.  But, the trigger detection rates remained above \%94 irrespective of fluency.  In fact, 3 out of 4 ASL users' trigger word is detected with 100\% accuracy.

 

\begin{table}[!t]
\centering
\caption{Comparison of DNNs for MDI Classification}
\begin{tabular}{>{\centering\arraybackslash}m{2cm}|>{\centering\arraybackslash}m{0.8cm}|>{\centering\arraybackslash}m{0.8cm}|>{\centering\arraybackslash}m{0.8cm}|>{\centering\arraybackslash}m{1.8cm} }
   \textbf{Architecture} & \textbf{$\mu$D} & \textbf{RD Map} & \textbf{RA Map} & \textbf{Feature-Level Fusion} \\ 
  \hline
 \hline
 CNN + BiLSTM &  78.8\% & 72.8\% & 67.5\% & 84.3 \% \\
 \hline
 CNN + BiLSTM + CTC & 80.6\% & 78.4\% & 71.3\% & 87.5\%\\
 \hline
 CNN + BiLSTM + CTC + MTL & \textbf{83.6\%} & \textbf{78.6\%} & \textbf{71.4\%} & \textbf{\textcolor{red}{JD-MIMTL 92\%}}\\
 \hline 
\end{tabular}
\label{tab:seq class table}
\end{table}

\subsection{Discussion}
Because RF sensors rely on kinetic properties of signing during recognition, signs that inherently contain greater movement (especially inter-sign movements) are easier to recognize.  For example, the signs \textsc{teacher} and \textsc{teach} both involve raising the hands to the level of the head, whereas \textsc{month} involves just a short swipe of a finger downward and \textsc{night} involves a more subtle downward, curved motion of the hand/arm, resulting in a detection rate that is over 20\% lower.  Effective ASL-based device triggering will require the design of a unique sign for this purpose, as commonly used daily expressions may mistakenly trigger a device.  In this regard, it is important to note that it is not necessary for such a trigger sign to have meaning in English; e.g. that \textsc{knock} might be sensible in meaning has little bearing on efficacy in terms of detectability, practical and cultural considerations.  In future work, we aim to work with deaf community partners to jointly evaluate usability and efficacy of kinetically unique trigger signs.

Another important consideration for device operation with ASL is real-time implementation on dedicated edge computing platforms.  Although there have been some studies of real-time gesture recognition using micro-Doppler signatures \cite{Sun2020,MGesture2021,Chmurski2021,Ninos2021}, these works have considered only a small number of classes (less than 12), and focus on hardware acceleration or reduction of the computational complexity of the model itself.  However, our initial work \cite{Oladipupo2021} in evaluating computational latency in the processing pipeline has shown that a significant part of the latency is not in the classification stage, but in the computation of the input representations themselves, especially micro-Doppler signatures. Latency depends not just on the duration (length) of the data, but also on short-time Fourier transform parameters, such as window length and overlap, which determine the dimensionality of the resulting spectrogram and impacts classification accuracy. Joint optimization of input representation generation and DNN model will be necessary to maximize real-time recognition performance.

\section{Conclusion}
\label{sec:Conc}
The proposed techniques in this paper enable trigger sign detection for device activation and sequential recognition of ASL in the context of daily living.
While conventional approaches to RF signal classification utilize just one RF data representation, this work exploits $\mu $D spectrograms, RD maps, and RA maps in a JD-MIMTL framework for sequential classification.  By defining tasks in terms of physically-relevant concepts for ASL recognition, sequences involving a mixture of 18 different daily activities and ASL signs was classified with 92\% accuracy.  The proposed double-thresholding trigger detection method achieves detection rates of 96\% and 98.9\% for non-ASL and ASL users, respectively, for the sign \textsc{teacher}.  Potential selections for trigger signs are evaluated based on sequential activity recognition accuracy and replicability across the fluency levels of users.  The results demonstrate the potential for RF sensing to be used for ASL-sensitive HCI.


\section*{Acknowledgment}
\label{sec:Acknow}
This work was funded in part by the National Science Foundation (NSF) Awards \#1932547, \#1931861, and \#1734938. Human studies research was conducted under UA Institutional Review Board (IRB) Protocol \#18-06-1271. 
\bibliographystyle{IEEEtran}
\bibliography{main}

\begin{thebibliography}{10}
\providecommand{\url}[1]{#1}
\csname url@samestyle\endcsname
\providecommand{\newblock}{\relax}
\providecommand{\bibinfo}[2]{#2}
\providecommand{\BIBentrySTDinterwordspacing}{\spaceskip=0pt\relax}
\providecommand{\BIBentryALTinterwordstretchfactor}{4}
\providecommand{\BIBentryALTinterwordspacing}{\spaceskip=\fontdimen2\font plus
\BIBentryALTinterwordstretchfactor\fontdimen3\font minus
  \fontdimen4\font\relax}
\providecommand{\BIBforeignlanguage}[2]{{%
\expandafter\ifx\csname l@#1\endcsname\relax
\typeout{** WARNING: IEEEtran.bst: No hyphenation pattern has been}%
\typeout{** loaded for the language `#1'. Using the pattern for}%
\typeout{** the default language instead.}%
\else
\language=\csname l@#1\endcsname
\fi
#2}}
\providecommand{\BIBdecl}{\relax}
\BIBdecl

\bibitem{Sun2016_IMU_sEMG}
J.~{Wu} \emph{et~al.}, ``A wearable system for recognizing american sign
  language in real-time using imu and surface emg sensors,'' \emph{IEEE Journal
  of Biomedical and Health Informatics}, vol.~20, no.~5, pp. 1281--1290, 2016.

\bibitem{Siddiqui2021}
N.~{Siddiqui} and R.~H.~M. {Chan}, ``Hand gesture recognition using multiple
  acoustic measurements at wrist,'' \emph{IEEE Transactions on Human-Machine
  Systems}, vol.~51, no.~1, pp. 56--62, 2021.

\bibitem{6814287}
M.~{Mohandes}, M.~{Deriche}, and J.~{Liu}, ``Image-based and sensor-based
  approaches to arabic sign language recognition,'' \emph{IEEE Transactions on
  Human-Machine Systems}, vol.~44, no.~4, pp. 551--557, 2014.

\bibitem{Koller2020_ASLvideo}
O.~{Koller}, N.~C. {Camgoz} \emph{et~al.}, ``Weakly supervised learning with
  multi-stream cnn-lstm-hmms to discover sequential parallelism in sign
  language videos,'' \emph{IEEE Transactions on Pattern Analysis and Machine
  Intelligence}, vol.~42, no.~9, pp. 2306--2320, 2020.

\bibitem{Cui2019_ASLvideo}
R.~{Cui}, H.~{Liu}, and C.~{Zhang}, ``A deep neural framework for continuous
  sign language recognition by iterative training,'' \emph{IEEE Transactions on
  Multimedia}, vol.~21, no.~7, pp. 1880--1891, 2019.

\bibitem{Sun2013_ASLkinect}
C.~{Sun}, T.~{Zhang}, B.~{Bao}, C.~{Xu}, and T.~{Mei}, ``Discriminative
  exemplar coding for sign language recognition with kinect,'' \emph{IEEE
  Transactions on Cybernetics}, vol.~43, no.~5, pp. 1418--1428, 2013.

\bibitem{Mittal2019_ASLleap}
A.~{Mittal}, P.~{Kumar}, P.~P. {Roy} \emph{et~al.}, ``A modified lstm model for
  continuous sign language recognition using leap motion,'' \emph{IEEE Sensors
  J.}, vol.~19, no.~16, pp. 7056--7063, 2019.

\bibitem{bragg2019sign}
D.~Bragg, O.~Koller, M.~Bellard, L.~Berke, P.~Boudreault, A.~Braffort,
  N.~Caselli, M.~Huenerfauth, H.~Kacorri, T.~Verhoef \emph{et~al.}, ``Sign
  language recognition, generation, and translation: An interdisciplinary
  perspective,'' in \emph{The 21st international ACM SIGACCESS conference on
  computers and accessibility}, 2019, pp. 16--31.

\bibitem{ASL_Patent2018}
S.~Gurbuz, A.~Gurbuz, C.~Crawford, and D.~Griffin, ``Radar-based methods and
  apparatus for communication and interpretation of sign languages,'' in
  \emph{U.S. Patent App. No. US2020/0334452 (Invention Disclosure filed Feb.
  2018; Prov. Patent App. filed Apr. 2019.)}, Oct. 2020.

\bibitem{Gurbuz_ASLR_RadarCon}
S.~Z. {Gurbuz}, A.~C. {Gurbuz}, E.~A. {Malaia}, D.~J. {Griffin}, C.~{Crawford},
  M.~M. {Rahman}, R.~{Aksu}, E.~{Kurtoglu} \emph{et~al.}, ``A linguistic
  perspective on radar micro-doppler analysis of american sign language,'' in
  \emph{2020 IEEE International Radar Conference (RADAR)}, 2020, pp. 232--237.

\bibitem{Gurbuz2020_ASL_SenConf}
S.~Z. {Gurbuz}, A.~C. {Gurbuz}, E.~A. {Malaia}, D.~J. {Griffin}, C.~{Crawford},
  E.~{Kurtoglu}, M.~M. {Rahman}, R.~{Aksu}, and R.~{Mdrafi}, ``Asl recognition
  based on kinematics derived from a multi-frequency rf sensor network,'' in
  \emph{2020 IEEE SENSORS}, 2020, pp. 1--4.

\bibitem{Gurbuz2021_ASLR}
S.~Z. {Gurbuz}, A.~C. {Gurbuz}, E.~A. {Malaia}, D.~J. {Griffin}, C.~S.
  {Crawford}, M.~M. {Rahman}, E.~{Kurtoglu} \emph{et~al.}, ``American sign
  language recognition using rf sensing,'' \emph{IEEE Sensors J.}, vol.~21,
  no.~3, pp. 3763--3775, 2021.

\bibitem{Gurbuz_DNN4Radar2020}
S.~Gurbuz, S.~Sun, and D.~Tahmoush, ``Radar systems, signals, and
  phenomenology,'' in \emph{Deep Neural Network Design for Radar Applications},
  S.~Gurbuz, Ed.\hskip 1em plus 0.5em minus 0.4em\relax IET, December 2020.

\bibitem{9106153}
V.~{Chen}, \emph{The Micro-Doppler Effect in Radar}, 2011.

\bibitem{1550191}
N.~Boulgouris, D.~Hatzinakos, and K.~Plataniotis, ``Gait recognition: a
  challenging signal processing technology for biometric identification,''
  \emph{IEEE Signal Processing Magazine}, vol.~22, no.~6, pp. 78--90, 2005.

\bibitem{8333730}
B.~Vandersmissen, Knudde \emph{et~al.}, ``Indoor person identification using a
  low-power fmcw radar,'' \emph{IEEE Transactions on Geoscience and Remote
  Sensing}, vol.~56, no.~7, pp. 3941--3952, 2018.

\bibitem{4801689}
Y.~Kim and H.~Ling, ``Human activity classification based on micro-doppler
  signatures using a support vector machine,'' \emph{IEEE Transactions on
  Geoscience and Remote Sensing}, vol.~47, no.~5, pp. 1328--1337, 2009.

\bibitem{szgMicroDoppler}
S.~Gurbuz, J.~Soraghan, A.~Balleri \emph{et~al.}, ``Micro-doppler based in-home
  aided and unaided walking recognition with multiple radar and sonar
  systems,'' \emph{IET Radar, Sonar \& Navigation}, vol.~11, 06 2016.

\bibitem{7426551}
M.~G. Amin, Y.~D. Zhang, F.~Ahmad, and K.~D. Ho, ``Radar signal processing for
  elderly fall detection: The future for in-home monitoring,'' \emph{IEEE
  Signal Processing Magazine}, vol.~33, no.~2, pp. 71--80, 2016.

\bibitem{6945894}
B.~Y. Su, K.~C. Ho, M.~J. Rantz, and M.~Skubic, ``Doppler radar fall activity
  detection using the wavelet transform,'' \emph{IEEE Transactions on
  Biomedical Engineering}, vol.~62, no.~3, pp. 865--875, 2015.

\bibitem{8613848}
A.-K. Seifert, M.~G. Amin \emph{et~al.}, ``Toward unobtrusive in-home gait
  analysis based on radar micro-doppler signatures,'' \emph{IEEE Transactions
  on Biomedical Engineering}, vol.~66, no.~9, pp. 2629--2640, 2019.

\bibitem{Gesture_Gu2019}
C.~Gu, J.~Wang, and J.~Lien, ``Motion sensing using radar: Gesture interaction
  and beyond,'' \emph{IEEE Microwave Magazine}, vol.~20, no.~8, pp. 44--57,
  2019.

\bibitem{GestureRadar2021}
Z.~{Wang}, Z.~{Yu}, X.~{Lou}, B.~{Guo}, and L.~{Chen}, ``Gesture-radar: A dual
  doppler radar based system for robust recognition and quantitative profiling
  of human gestures,'' \emph{IEEE Transactions on Human-Machine Systems},
  vol.~51, no.~1, pp. 32--43, 2021.

\bibitem{Arbabian2012}
A.~{Arbabian} \emph{et~al.}, ``A 94ghz mm-wave to baseband pulsed-radar for
  imaging and gesture recognition,'' in \emph{Symp. VLSI Cir.}, 2012, pp.
  56--57.

\bibitem{Gurbuz_SPM_2019}
S.~Z. {Gurbuz} and M.~G. {Amin}, ``Radar-based human-motion recognition with
  deep learning: Promising applications for indoor monitoring,'' \emph{IEEE
  Signal Processing Magazine}, vol.~36, no.~4, pp. 16--28, 2019.

\bibitem{Wang_SGRUN2018}
\BIBentryALTinterwordspacing
M.~Wang, G.~Cui, X.~Yang, and L.~Kong, ``Human body and limb motion recognition
  via stacked gated recurrent units network,'' \emph{IET Radar, Sonar \&
  Navigation}, vol.~12, no.~9, pp. 1046--1051, 2018. [Online]. Available:
  \url{https://ietresearch.onlinelibrary.wiley.com/doi/abs/10.1049/iet-rsn.2018.5054}
\BIBentrySTDinterwordspacing

\bibitem{Latern2018}
Z.~{Zhang}, Z.~{Tian}, and M.~{Zhou}, ``Latern: Dynamic continuous hand gesture
  recognition using fmcw radar sensor,'' \emph{IEEE Sensors Journal}, vol.~18,
  no.~8, pp. 3278--3289, 2018.

\bibitem{Soli2016}
J.~Lien \emph{et~al.}, ``Soli: Ubiquitous gesture sensing with millimeter wave
  radar,'' \emph{ACM Trans. Graph.}, vol.~35, no.~4, Jul. 2016.

\bibitem{Santra2018}
S.~{Hazra} and A.~{Santra}, ``Robust gesture recognition using millimetric-wave
  radar system,'' \emph{IEEE Sensors Letters}, vol.~2, no.~4, pp. 1--4, 2018.

\bibitem{Santra2019}
S.~Hazra and A.~Santra, ``Short-range radar-based gesture recognition system
  using 3d cnn with triplet loss,'' \emph{IEEE Access}, vol.~7, pp.
  125\,623--125\,633, 2019.

\bibitem{mmASL}
\BIBentryALTinterwordspacing
P.~S. Santhalingam, A.~A. Hosain, D.~Zhang, P.~Pathak, H.~Rangwala, and
  R.~Kushalnagar, ``Mmasl: Environment-independent asl gesture recognition
  using 60 ghz millimeter-wave signals,'' \emph{Proc. ACM Interact. Mob.
  Wearable Ubiquitous Technol.}, vol.~4, no.~1, Mar. 2020. [Online]. Available:
  \url{https://doi.org/10.1145/3381010}
\BIBentrySTDinterwordspacing

\bibitem{Gu2019}
Z.~{Gu} \emph{et~al.}, ``Blind separation of doppler human gesture signals
  based on continuous-wave radar sensors,'' \emph{IEEE Transactions on
  Instrumentation and Measurement}, vol.~68, no.~7, pp. 2659--2661, 2019.

\bibitem{Haobo2021}
H.~Li, A.~Mehul, J.~Le~Kernec, S.~Z. Gurbuz, and F.~Fioranelli, ``Sequential
  human gait classification with distributed radar sensor fusion,'' \emph{IEEE
  Sensors Journal}, vol.~21, no.~6, pp. 7590--7603, 2021.

\bibitem{Ding2019}
C.~Ding, H.~Hong, Y.~Zou, H.~Chu, X.~Zhu, F.~Fioranelli, J.~Le~Kernec, and
  C.~Li, ``Continuous human motion recognition with a dynamic range-doppler
  trajectory method based on fmcw radar,'' \emph{IEEE Transactions on
  Geoscience and Remote Sensing}, vol.~57, no.~9, pp. 6821--6831, 2019.

\bibitem{Kulhandjian2019}
H.~Kulhandjian, P.~Sharma, M.~Kulhandjian, and C.~D'Amours, ``Sign language
  gesture recognition using doppler radar and deep learning,'' in \emph{2019
  IEEE Globecom Workshops (GC Wkshps)}, 2019, pp. 1--6.

\bibitem{Gurbuz_JSEN_2021}
S.~Z. Gurbuz, M.~Mahbubur~Rahman, E.~Kurtoglu, E.~Malaia, A.~C. Gurbuz, D.~J.
  Griffin, and C.~Crawford, ``Multi-frequency rf sensor fusion for word-level
  fluent asl recognition,'' \emph{IEEE Sensors Journal}, pp. 1--1, 2021.

\bibitem{Mahbub_TAES2021}
M.~Mahbubur~Rahman, E.~Malaia, A.~C. Gurbuz, D.~J. Griffin, C.~Crawford, and
  S.~Gurbuz, ``Effect of kinematics and fluency in adversarial synthetic data
  generation for asl recognition with rf sensors,'' \emph{IEEE Transactions on
  Aerospace and Electronics Systems}, pp. 1--1, 2021.

\bibitem{Melgarejo2014}
P.~Melgarejo \emph{et~al.}, ``Leveraging directional antenna capabilities for
  fine-grained gesture recognition,'' in \emph{Proc. ACM UbiComp.}, 2014, p.
  541–551.

\bibitem{WiFinger2016}
H.~Li \emph{et~al.}, ``Wifinger: Talk to your smart devices with finger-grained
  gesture,'' in \emph{Proc. ACM UbiComp}, 2016, p. 250–261.

\bibitem{malaia2016assessment}
E.~Malaia, J.~D. Borneman, and R.~B. Wilbur, ``Assessment of information
  content in visual signal: analysis of optical flow fractal complexity,''
  \emph{Visual Cognition}, vol.~24, no.~3, pp. 246--251, 2016.

\bibitem{borneman2018motion}
J.~D. Borneman, E.~Malaia, and R.~B. Wilbur, ``Motion characterization using
  optical flow and fractal complexity,'' \emph{Journal of Electronic Imaging},
  vol.~27, no.~5, p. 051229, 2018.

\bibitem{malaia2020syllable}
E.~A. Malaia and R.~B. Wilbur, ``Syllable as a unit of information transfer in
  linguistic communication: The entropy syllable parsing model,'' \emph{Wiley
  Interdisciplinary Reviews: Cognitive Sci.}, vol.~11, no.~1, p. e1518, 2020.

\bibitem{blumenthal2019shared}
A.~Blumenthal-Dram{\'e} and E.~Malaia, ``Shared neural and cognitive mechanisms
  in action and language: The multiscale information transfer framework,''
  \emph{Wiley Interdisciplinary Reviews: Cognitive Science}, vol.~10, no.~2, p.
  e1484, 2019.

\bibitem{wilbur2008contributions}
R.~B. Wilbur and E.~Malaia, ``Contributions of sign language research to
  gesture understanding: What can multimodal computational systems learn from
  sign language research,'' \emph{International journal of semantic computing},
  vol.~2, no.~01, pp. 5--19, 2008.

\bibitem{DeepASL2017}
B.~Fang, J.~Co, and M.~Zhang, ``Deepasl: Enabling ubiquitous and non-intrusive
  word and sentence-level sign language translation,'' in \emph{Proceedings of
  the 15th ACM Conference on Embedded Network Sensor Systems}.\hskip 1em plus
  0.5em minus 0.4em\relax Association for Computing Machinery, 2017.

\bibitem{SignFi2018}
Y.~Ma, G.~Zhou, S.~Wang, H.~Zhao, and W.~Jung, ``Signfi: Sign language
  recognition using wifi,'' \emph{Proc. ACM Interact. Mob. Wearable Ubiquitous
  Technol.}, vol.~2, no.~1, Mar. 2018.

\bibitem{beal2019hearing}
J.~S. Beal and K.~Faniel, ``Hearing l2 sign language learners,'' \emph{Sign
  Language Studies}, vol.~19, no.~2, pp. 204--224, 2019.

\bibitem{9425571}
S.~Z. Gurbuz, M.~Mahbubur~Rahman, E.~Kurtoglu \emph{et~al.}, ``Multi-frequency
  rf sensor fusion for word-level fluent asl recognition,'' \emph{IEEE Sensors
  Journal}, pp. 1--1, 2021.

\bibitem{Caselli2017}
N.~K. Caselli, Z.~S. Sehyr, A.~M. Cohen-Goldberg, and K.~Emmorey, ``Asl-lex: A
  lexical database of american sign language,'' \emph{Behavior Research
  Methods}, vol.~49, no.~2, pp. 784--801, Apr 2017.

\bibitem{10.5555/888857}
B.~K. Horn and B.~G. Schunck, ``Determining optical flow,'' USA, Tech. Rep.,
  1980.

\bibitem{Hill2012_BlackASL}
J.~Hill, ``Black asl,'' \emph{Journal of American Sign Languages and
  Literatures}.

\bibitem{4469865}
P.~V. {Dorp} and F.~C.~A. {Groen}, ``Feature-based human motion parameter
  estimation with radar,'' \emph{IET Radar, Sonar Navigation}, vol.~2, no.~2,
  pp. 135--145, 2008.

\bibitem{7165625}
C.~{Karabacak}, S.~Z. {Gurbuz}, A.~C. {Gurbuz} \emph{et~al.}, ``Knowledge
  exploitation for human micro-doppler classification,'' \emph{IEEE Geoscience
  and Remote Sensing Letters}, vol.~12, no.~10, pp. 2125--2129, 2015.

\bibitem{dfd}
T.~Eiter and H.~Mannila, ``Computing discrete frechet distance,'' Christian
  Doppler Lab., Vienna Univ. of Technology, Tech. Rep. 94/64, 1994.

\bibitem{optmatch}
A.~ABBOTT and A.~TSAY, ``Sequence analysis and optimal matching methods in
  sociology: Review and prospect,'' \emph{Sociological Methods \& Research},
  vol.~29, no.~1, pp. 3--33, 2000.

\bibitem{8188276}
Y.~{Vaezi} and M.~{Van der Baan}, ``Comparison of the sta/lta and power
  spectral density methods for microseismic event detection,''
  \emph{Geophysical Journal International}, vol. 203, no.~3, pp. 1896--1908,
  2015.

\bibitem{Sun2020}
Y.~{Sun} \emph{et~al.}, ``Real-time radar-based gesture detection and
  recognition built in an edge-computing platform,'' \emph{IEEE Sensors
  Journal}, vol.~20, no.~18, pp. 10\,706--10\,716, 2020.

\bibitem{9531622}
Y.~Sun, H.~Xiong, D.~K.~P. Tan, T.~X. Han, R.~Du, X.~Yang, and T.~T. Ye,
  ``Moving target localization and activity/gesture recognition for indoor
  radio frequency sensing applications,'' \emph{IEEE Sensors Journal}, pp.
  1--1, 2021.

\bibitem{rs12030454}
Z.~Zeng, M.~G. Amin, and T.~Shan, ``Arm motion classification using time-series
  analysis of the spectrogram frequency envelopes,'' \emph{Remote Sensing},
  vol.~12, no.~3, 2020.

\bibitem{seqlen}
F.~Jafariakinabad, S.~Tarnpradab, and K.~Hua, ``Syntactic recurrent neural
  network for authorship attribution,'' 02 2019.

\bibitem{DBLP:journals/corr/abs-1708-02709}
T.~Young, D.~Hazarika, S.~Poria, and E.~Cambria, ``Recent trends in deep
  learning based natural language processing,'' \emph{CoRR}, vol.
  abs/1708.02709, 2017.

\bibitem{6638947}
A.~{Graves}, A.~{Mohamed}, and G.~{Hinton}, ``Speech recognition with deep
  recurrent neural networks,'' in \emph{2013 IEEE International Conference on
  Acoustics, Speech and Signal Processing}, 2013, pp. 6645--6649.

\bibitem{multitask}
R.~Caruana, ``Multitask learning,'' \emph{Machine Learning}, vol.~28, 07 1997.

\bibitem{Graves06connectionisttemporal}
A.~Graves, S.~Fernández, and F.~Gomez, ``Connectionist temporal
  classification: Labelling unsegmented sequence data with recurrent neural
  networks,'' in \emph{in Proc. Int. Conf. on Mach. Learn.}, 2006, pp.
  369--376.

\bibitem{7410526}
R.~{Girshick}, ``Fast r-cnn,'' in \emph{Proc. IEEE ICCV}, 2015, pp. 1440--1448.

\bibitem{10.1145/1390156.1390177}
R.~Collobert and J.~Weston, ``A unified architecture for natural language
  processing: Deep neural networks with multitask learning,'' ser. ICML
  '08.\hskip 1em plus 0.5em minus 0.4em\relax New York, NY, USA: Association
  for Computing Machinery, 2008, p. 160–167.

\bibitem{8848395}
T.~{Gong}, T.~{Lee}, C.~{Stephenson}, V.~{Renduchintala}, S.~{Padhy},
  A.~{Ndirango}, G.~{Keskin}, and O.~H. {Elibol}, ``A comparison of loss
  weighting strategies for multi task learning in deep neural networks,''
  \emph{IEEE Access}, vol.~7, pp. 141\,627--141\,632, 2019.

\bibitem{8578879}
R.~{Cipolla}, Y.~{Gal}, and A.~{Kendall}, ``Multi-task learning using
  uncertainty to weigh losses for scene geometry and semantics,'' in
  \emph{Proc. IEEE/CVF Conf. on Comp. Vis. and Patt. Recog.}, 2018, pp.
  7482--7491.

\bibitem{siri}
``Hey siri: An on-device dnn-powered voice trigger for apple's personal
  assistant.''\hskip 1em plus 0.5em minus 0.4em\relax Apple Machine Learning
  Research, October 2017.

\bibitem{malaia2017current}
E.~Malaia, ``Current and future methodologies for quantitative analysis of
  information transfer in sign language and gesture data,'' \emph{Behavioral
  and Brain Sciences}, vol.~40, 2017.

\bibitem{ford2021classification}
L.~K. Ford, J.~D. Borneman, J.~Krebs, E.~A. Malaia, and B.~P. Ames,
  ``Classification of visual comprehension based on eeg data using sparse
  optimal scoring,'' \emph{J. Neural Engineering}, vol.~18, no.~2, p. 026025,
  2021.

\bibitem{9027708}
T.~{Tsai} and P.~{Hao}, ``Customized wake-up word with key word spotting using
  convolutional neural network,'' in \emph{2019 International SoC Design
  Conference (ISOCC)}, 2019, pp. 136--137.

\bibitem{MGesture2021}
H.~Liu, A.~Zhou, Z.~Dong, Y.~Sun, J.~Zhang, L.~Liu, H.~Ma, J.~Liu, and N.~Yang,
  ``M-gesture: Person-independent real-time in-air gesture recognition using
  commodity millimeter wave radar,'' \emph{IEEE Internet of Things Journal},
  pp. 1--1, 2021.

\bibitem{Chmurski2021}
M.~Chmurski, M.~Zubert, K.~Bierzynski, and A.~Santra, ``Analysis of
  edge-optimized deep learning classifiers for radar-based gesture
  recognition,'' \emph{IEEE Access}, vol.~9, pp. 74\,406--74\,421, 2021.

\bibitem{Ninos2021}
A.~Ninos, J.~Hasch, and T.~Zwick, ``Real-time macro gesture recognition using
  efficient empirical feature extraction with millimeter-wave technology,''
  \emph{IEEE Sensors Journal}, vol.~21, no.~13, pp. 15\,161--15\,170, 2021.

\bibitem{Oladipupo2021}
\BIBentryALTinterwordspacing
O.~O. Adeoluwa, S.~J. Kearney, E.~Kurtoglu, C.~J. Connors, and S.~Z. Gurbuz,
  ``{Near real-time ASL recognition using a millimeter wave radar},'' in
  \emph{Radar Sensor Technology XXV}, K.~I. Ranney and A.~M. Raynal, Eds., vol.
  11742, International Society for Optics and Photonics.\hskip 1em plus 0.5em
  minus 0.4em\relax SPIE, 2021, pp. 173 -- 184. [Online]. Available:
  \url{https://doi.org/10.1117/12.2588616}
\BIBentrySTDinterwordspacing

\end{thebibliography}



\end{document}